\documentclass[lettersize,journal]{IEEEtran}
\usepackage{amsmath,amsfonts}
\usepackage{algorithmic}
\usepackage{algorithm}
\usepackage{array}
\usepackage[caption=false,font=normalsize,labelfont=sf,textfont=sf]{subfig}
\usepackage{textcomp}
\usepackage{stfloats}
\usepackage{url}
\usepackage{verbatim}
\usepackage{graphicx}
\usepackage{cite}
\usepackage{adjustbox}

\usepackage{diagbox}

\usepackage{booktabs} % for creating beautiful horizontal lines
\usepackage{multirow} % for merging table rows

\usepackage{hyperref}

\hypersetup{hidelinks,
	colorlinks=true,
	allcolors=blue,
	pdfstartview=Fit,
	breaklinks=true}

\usepackage{makecell}
\usepackage{bm}
\usepackage{amssymb}
\usepackage{multirow}
\usepackage[table,xcdraw]{xcolor}
\usepackage[T1]{fontenc}
\usepackage{array} 
\usepackage{booktabs}
\usepackage[justification=centering]{caption}

\begin{document}

\title{\fontsize{22}{26}\selectfont A Model Consistency-Based Countermeasure to GAN-Based Data Poisoning Attack in Federated Learning}

\author{Wei Sun, Bo Gao, \IEEEmembership{Member, IEEE}, Ke Xiong, \IEEEmembership{Member, IEEE}, Yuwei Wang \\ Pingyi Fan, \IEEEmembership{Senior Member, IEEE}, Khaled Ben Letaief, \IEEEmembership{Fellow, IEEE}

	\thanks{W. Sun, B. Gao and K. Xiong are with the Engineering Research Center of Network Management Technology for High Speed Railway of Ministry of Education, School of Computer Science and Technology, and the Collaborative Innovation Center of Railway Traffic Safety, Beijing Jiaotong University, Beijing 100044, China. E-mail: \{21120398, bogao, kxiong\}@bjtu.edu.cn.}
	\thanks{Y. Wang is with the Institute of Computing Technology, Chinese Academy of Sciences, Beijing 100190, China. E-mail: ywwang@ict.ac.cn.}
	\thanks{P. Fan is with the Beijing National Research Center for Information Science and Technology, and the Department of Electronic Engineering, Tsinghua University, Beijing 100084, China. E-mail: fpy@tsinghua.edu.cn.}
	\thanks{K. Letaief is with the Department of Electrical and Computer Engineering, Hong Kong University of Science and Technology, Hong Kong 999077, China. E-mail: eekhaled@ust.hk.}}

%\author{IEEE Publication Technology,~\IEEEmembership{Staff,~IEEE,}
        % <-this % stops a space
%\thanks{This paper was produced by the IEEE Publication Technology Group. They are in Piscataway, NJ.}% <-this % stops a space
%\thanks{Manuscript received April 19, 2021; revised August 16, 2021.}}

% The paper headers
%\markboth{Journal of \LaTeX\ Class Files,~Vol.~14, No.~8, August~2021}%
%{Shell \MakeLowercase{\textit{et al.}}: A Sample Article Using IEEEtran.cls for IEEE Journals}

%\IEEEpubid{0000--0000/00\$00.00~\copyright~2021 IEEE}
% Remember, if you use this you must call \IEEEpubidadjcol in the second
% column for its text to clear the IEEEpubid mark.

\maketitle

\begin{abstract}
In federated learning (FL), although the original intention of “available but not visible” data is to allay data privacy concerns, it potentially brings new security threats, particularly poisoning attacks that target such “not visible” local data. Intuitively, such data poisoning attacks have great potential in stealthily degrading global FL outcomes, and are expected to be even stealthier if being enhanced by generative models like generative adversarial networks (GANs). However, existing defense methods have not been thoroughly challenged in this regard and generally fail to be aware of a local generation of seemingly legitimate poisoned data. With a growing concern on potentially stealthier attacks, in this paper, a cost-effective defense mechanism named Model Consistency-Based Defense (MCD) is proposed, which offers a comprehensive examination of available local models across multiple feature dimensions, providing an indirect yet effective means of identifying hidden data poisoning attackers. To push the limit of MCD against stealthier attacks, we propose a new GAN-based data poisoning attack model named VagueGAN and an unsupervised variant of it, which can be flexibly deployed to generate seemingly legitimate but noisy poisoned data. The consistency of GAN outputs revealed by VagueGAN helps strengthen MCD to work against stealthier GAN-based attacks as well as other mainstream ones. Extensive experiments on multiple open datasets (MNIST, Fashion-MNIST, CIFAR-10, CIFAR-100, and Mini-Imagenet) indicate that our attack method better balances the trade-off between attack effectiveness and stealthiness with low complexity.  More importantly, our defense mechanism is shown to be more competent in identifying a variety of poisoned data, particularly stealthier GAN-poisoned ones.

%As a distributed machine learning paradigm, federated learning (FL) is collaboratively carried out on privately owned datasets but without direct data access. Although the original intention is to allay data privacy concerns, "available but not visible" data in FL potentially bring new security threats, particularly poisoning attacks that target such "not visible" local data. Initial attempts have been made to conduct data poisoning attacks against FL systems, but they cannot be fully successful due to their high chance of causing statistical anomalies. To unleash the potential for truly "invisible" attacks, in this paper, a new data poisoning attack model named VagueGAN is proposed, which can generate seemingly legitimate but noisy poisoned data by untraditionally taking advantage of generative adversarial network (GAN) variants. Capable of manipulating the quality of poisoned data on demand, VagueGAN enables to trade-off attack effectiveness and stealthiness. Furthermore, a countermeasure named Model Consistency-Based Defense (MCD) to this type of data poisoning attacks is proposed, which can identify GAN-poisoned data or models owing to the consistency of GAN outputs. Extensive experiments on multiple datasets indicate that our attack method is generally much more stealthy as well as more effective in degrading FL performance with low complexity. Our defense method is also shown to be more competent in identifying GAN-poisoned data or models. 
\end{abstract}

\begin{IEEEkeywords}
Fedrated Learning, Security and Privacy, Generative Adversarial Networks, Data Poisoning 
\end{IEEEkeywords}

\section{Introduction}
%Deep learning (DL) has been delivering promising results across a wide range of industries and applications. However, as the needs for data and computing power continue to grow, it has been increasingly costly and inefficient to centralize those resources at scale to support traditional deep learning in a centralized manner. Recently, distributed approaches to deep learning have drawn great attention, and are expected to play a key role in the upcoming 6G era toward connected intelligence~\cite{chen2020connected}.

Emerging as a promising distributed learning paradigm, federated learning (FL) can be used to collaboratively train a deep learning model at a server based on decentralized datasets privately owned by multiple clients. However, the ``available but not visible'' nature of training data in FL still leads to security risks. A server usually has to rely on the support of a large number of possibly untrustworthy clients, because it is very likely that some of the clients especially crowdsourced ones are originally manipulated or hijacked by some attackers. More seriously, the server does not have access to the clients' private datasets that largely determine the quality of training outcomes, so the local data ``not visible'' to the server or the resulting local models can easily become the best targets of the attackers~\cite{wei2024trustworthy}.

Poisoning attacks are the greatest threats in an FL system. An attacker can falsify either the local data (via data poisoning) or the local models (via model poisoning) of compromised clients to indirectly mislead the global model of a server that is built upon those local data or models. Theoretically, such poisoning attacks can be not only effective in undermining the global model but also undetectable by the server without data access. In comparison with the model poisoning attacks, the data poisoning attacks are generally harder to be detected by the server and thus would be more worth studying from both attack and defense perspectives ~\cite{rodriguez2023survey}.

However, existing research on stealthy data poisoning in FL remains in its infancy. Existing defense methods have not been really challenged due to lack of data poisoning attacks that can ensure both attack effectiveness and stealthiness~\cite{tolpegin2020data,upreti2022defending,shen2016auror,li2021lomar,sikandar2023detailed,zhang2024visualizing}. Existing attack methods, however, focus more on attack effectiveness rather than attack stealthiness, thus inevitably giving rise to noticeable changes in local model distributions that can be readily detected by existing defense methods~\cite{lyu2020threats, gosselin2022privacy,mothukuri2021survey,jere2020taxonomy,xia2023poisoning,tolpegin2020data,zhang2020poisongan,zhang2019poisoning,cao2019understanding}. Although the stealthiness of data poisoning attacks in FL has not been previously achieved in the existing work, such ``not visible'' attacks still show great potential in globally degrading FL outcomes while locally hiding themselves, so we must get fully prepared to defend FL systems against upcoming stealthier attacks.

In this paper, we aim to develop a countermeasure to stealthier data poisoning attacks in FL. We are particularly interested in thwarting the potential of generative adversarial network (GAN) models from achieving stealthier attacks. Targeting GAN-based data poisoning attacks as well as other mainstream ones, we propose a cost-effective defense method that captures the historical characteristics of available local models so as to indirectly identify hidden attackers. To push the limit of our defense method against stealthier attacks, we propose a new GAN-based data poisoning method with enhanced attack stealthiness~\cite{goodfellow2020generative}. Instead of producing near-realistic data of high quality through regular uses of GAN, our attack method turns to reversely leverage the power of GAN to generate vague data with appropriate amounts of poisonous noise. The local models poisoned by such vague data can be made globally damaging, while the vague data can keep the statistical characteristics of their GAN inputs so that the local models poisoned can be seemingly legitimate. In addition, it can be computationally efficient to generate vague data of relatively low quality, rendering data poisoning broadly applicable, even for resource-constrained mobile devices. Fortunately, our attack model reveals the consistency of GAN outputs and thus helps strengthen our defense method to work against stealthier GAN-based attacks as well as other mainstream ones. To more effectively establish a safer and more robust FL system, our contributions are as follows. 

\begin{itemize}
	\item In response to stealthier data poisoning attacks, we propose a cost-effective countermeasure named $Model$ $Consistency$ $Based$ $Defense$ $(MCD)$, which comprehensively reviews available local models, and identifies abnormal clients via two statistical metrics inspired by our findings on the consistency of GAN outputs.
	\item Before that, we propose an attack method named $VagueGAN$, which is a GAN model specially designed for stealthier data poisoning attacks against FL systems. It generates vague poisoned data that can not only effectively but also unnoticeably attack a global model with all labels/classes involved. \textit{To the best of our knowledge, our MCD is the first countermeasure to such a GAN-based attack emphasizing improved attack stealthiness}.
	\item We develop guidelines for taking full advantage of VagueGAN to achieve a balanced trade-off between attack effectiveness and stealthiness, which is not readily achievable for existing approaches. Furthermore, we propose an unsupervised variant of VagueGAN, along with flexible deployment options, making it broader applicable.
	\item We conduct extensive experiments on multiple most-used open datasets: i.e., MNIST, Fashion-MNIST, CIFAR-10, CIFAR-100, and Mini-Imagenet. It is verified that the data poisoning attacks enhanced by our VagueGAN not only better degrade FL outcomes with low efforts but also are generally much less detectable. More importantly, our MCD better identifies a variety of poisoned data, particularly stealthier GAN-poisoned ones.
	%\item We conduct extensive experiments on multiple mostused datasets: CIFAR-10, MNIST and Fashion-MNIST. It is verified that data poisoning attacks enhanced by our VagueGAN not only better degrade FL outcomes with low efforts but also are generally much less detectable. More importantly, regarding the defense against data poisoning attacks, our MCD better identifies a variety of poisoned data, particularly GAN-poisoned ones beyond traditional threats.
\end{itemize}

The remainder of this paper is organized as follows. Related work is discussed in section II. Objectives and assumptions are outlined in section III, where the system model and attack model are defined. Our proposed attack is elaborated in section IV. Our proposed defense is suggested in section V. Performance evaluation is presented in section VI. Finally, our conclusions and future work are summarized in section VII.

\section{Related Work}

Although data poisoning attacks and their countermeasures in regular deep learning systems have been well studied~\cite{hitaj2017deep,shi2018spectrum,huang2021data,zhang2020online}, the research on those in FL systems is still in its infancy.

To mitigate data poisoning attacks against FL systems, some targeted countermeasures have been found to be useful. Some existing works, e.g.,~\cite{tolpegin2020data,upreti2022defending,shejwalkar2021manipulating,zhang2022fldetector,shen2023privacy,awan2021contra,li2021lomar,mu2024feddmc} employ similar two-step statistical approaches, where lower-dimensional local models are obtained in the first step and are statistically analyzed for outlier detection in the second step. Specifically, in~\cite{tolpegin2020data} and~\cite{upreti2022defending}, the authors propose to use principal component analysis (PCA) and uniform manifold approximation and projection (UMAP) respectively to reduce the dimensions of all local models to two-dimensional, to facilitate the identification of any abnormal model distribution. In ~\cite{shejwalkar2021manipulating}, the authors propose proposed divide-and-conquer (DnC), which computes the distance between the maximum variance direction of all local models and the updated scalar product to identify anomalies. In~\cite{zhang2022fldetector}, the authors propose FLDetector, where the server predicts every client's model updates and flags a client as malicious if there is a multi-round mismatch between the client's uploaded model updates and its predicted ones. In~\cite{shen2023privacy}, the authors propose a label-flipping-robust (LFR) algorithm based on cosine similarity temporal analysis, which is applicable to both independent and identically distributed (IID) and non-IID data. In ~\cite{awan2021contra}, the authors propose CONTRA, which implements a cosine-similarity-based measure for outlier detection. In~\cite{li2021lomar}, the authors propose proposed local malicious factor (LoMar), which is based on the kernel density estimation and K-nearest neighbor methods. In~\cite{mu2024feddmc} , the authors propose FedDMC, which achieves clustering-based malicious client detection by applying a binary tree-based noise module to the dimension-reduced model parameters. In addition, some defense methods~\cite{wang2020model,cao2021provably,zhao2022detecting,li2019abnormal,10054157,ovi2023confident,lai2023two} detect malicious clients by leveraging additional resources or tools to characterize and identify abnormal local models or datasets. However, none of these defense methods works well if attackers can generate seemingly legitimate poisoned data whose statistical distributions approximate normal ones.

As for data poisoning attacks against FL systems, existing works primarily leverage label flipping techniques to generate poisoned local data through falsifying certain labels of training data. Specifically, in~\cite{tolpegin2020data}, the authors propose a label flipping attack to rearrange the label-sample associations of a local dataset to generate a poisoned dataset. In addition, in~\cite{hitaj2017deep}, the authors propose a GAN model for the first time to extract private features of local data from GAN-generated pseudo data after initializing the GAN by the global model. On this basis, in~\cite{zhang2020poisongan}, the authors propose to enhance regular label flipping attacks by their PoisonGAN, an off-the-shelf GAN model only used to enlarge the sizes of local datasets with legitimate pseudo samples by using a discriminator initialized by the global model. In~\cite{yang2023clean}, the authors propose a clean-label poisoning attack by adding a perturbation to the gradients of adversarial loss, which only generates poisoned features without falsifying labels. However, these attack methods inevitably give rise to significant changes in the local distributions of training data, which can be readily detected by corresponding defense methods. Consequently, they do not currently represent a significant threat, therefore cannot be used to cultiave a robust defense system.

%Article~\cite{lai2023two} proposes DPA-FL, which tests the client model with a separate test dataset and finds the attackers when its accuracy is low.

%In short, the above defense methods are all carried out in two steps. The first step is to reduce the dimensionality of every high-dimensional model, and the PCA is the most commonly used dimensionality reduction method. The second step is to statistically detect models with large outliers. 

\begin{figure*}[h]
	\centering
	\includegraphics[width=\linewidth]{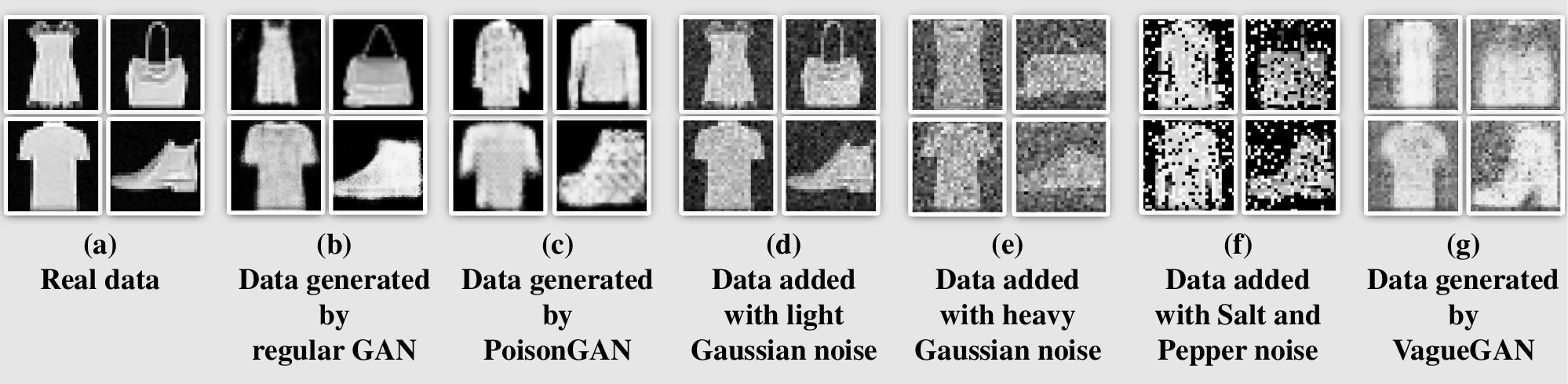}
	\caption{Comparison of GAN-generated and noise-superimposed samples: (a) Real data; (b) Data generated by regular GAN; (c) Data generated by PoisonGAN for data augmentation only; (d) Data added with light Gaussian noise; (e) Data added with heavy Gaussian noise; (f) Data added with Salt and Pepper noise; (g) Data generated by VagueGAN for direct attack use.}
\end{figure*}

\section{Objectives and Assumptions}

In this section, we describe the objectives and assumptions in an FL system.  

\subsection{System Model}

We consider an FL system that consists of a server and a set $\mathcal C = \left\{c_1,c_2,...,c_N \right\}$ of $N$ mobile clients. A regular training process of FL is considered. Initially, the server distributes an FL main task to each client. In each round $t = 1,2,...,T$ of FL, every client $c_i$ receives model updates from the server and trains a local model $\bm{\theta}_{t,i}$ on its local dataset $\mathcal D^{train}_i$. To avoid potential network congestions, the server then randomly selects $K$ out of $N$ clients, aggregates their local models into a global model $\bm{\tilde{\theta}}_{t}$ and sends it to all the clients for a new round of training. Meanwhile, the server uses a test dataset $\mathcal D^{test}$ to obtain the global accuracy $a_t$ of $\bm{\tilde{\theta}}_{t}$. %Table I summarizes the notation utilized in the described FL system.

\subsection{Attack Model}

\paragraph{\textbf{Attacker's Goal}} In an FL system, we assume that there exists at least one data poisoning attacker. The attacker's primary goal is to degrade the performance of the global model as much as possible. Second, the attacker's another equally important goal is to hide its attacks from being detected by the server.

%\begin{figure}[htbp]
%	\centering
%	\includegraphics[width=\linewidth]{fig1}
%	\caption{Basic system model of FL system.}
%	\label{fig1}
%\end{figure}

\paragraph{\textbf{Attacker's Capabilities}} The attacker is assumed to be capable of carrying out data poisoning attacks. First, the attacker can gain control of one or more benign clients or disguise itself as a benign client. Second, the attacker can gain all the privileges of each controlled client, e.g., local dataset, local model training process, etc. 

Due to our focus on data poisoning attacks, the attacker’s scope of operation is assumed to be limited to every controlled client. First, we guarantee that the server is completely honest and unattackable. Second, we assume that communication links for FL are reliable and the attacker cannot influence the process of model exchanges over the links. Third, the attacker cannot access any datasets of the server and uncontrolled clients. Finally, the attacker also has no access to the models of uncontrolled clients. 

\paragraph{\textbf{Attacker's Approach}} The attacker carries out a data poisoning attack to achieve its goals. In particular, the attacker first obtains control of at least one client as malicious client through certain means. Then the attacker poisons the local dataset of the malicious client and obtains a poisoned local dataset using an attack method. Finally, the malicious client trains a poisoned local model on its poisoned local dataset and uploads it to the server. If not being detected by the server, the poisoned local model is expected to harm the global model after model aggregation. 

%\begin{figure}[htbp]
%	\centering{\includegraphics{model.png}}
%	\caption{General idea of a data poisoning attack.}
%	\label{fig1}
%\end{figure}

\section{GAN-Based Data Poisoning Attacks }

In this section, to push the limit of our defense method against stealthier attacks, we propose a new GAN model called VagueGAN, which is specifically designed for stesalthier data poisoning attacks against FL systems. VagueGAN can help strengthen our defense method to work against GAN-based attacks as well as other mainstream ones. 

To this end, subsection IV-A introduces the architecture, loss function, and other design considerations underlying VagueGAN's poisoning ability. Subsection IV-B outlines the overall workflow of a data poisoning attack enabled by VagueGAN. To provide deeper insights, subsection IV-C presents a comprehensive theoretical analysis. Subsection IV-D gives practical guidelines for implementing VagueGAN for a trade-off between attack effectiveness and stealthiness. Finally, subsection IV-E introduces an unsupervised variant of VagueGAN, while subsection IV-F discusses deployment advantages of VagueGAN. 

\subsection{VagueGAN Model}

A GAN model has a generator $G$ and a discriminator $D$. On the one hand, the generator $G$ generates fake data samples close to real ones in terms of  distribution and tries to convince the discriminator $D$. On the other hand, the discriminator $D$ tries to distinguish between real and fake samples. They compete with each other in the training phase, thus operating as an adversarial game that makes the generator $G$ outputs approximate original data in distribution.

Our VagueGAN unconventionally leverages the power of GAN to generate seemingly legitimate vague data with appropriate amounts of poisonous noise, in order to achieve a balanced trade-off between attack effectiveness and stealthiness. On the one hand, VagueGAN needs to generate seemingly legitimate poisoned data for a stealthier attack. We extend the Conditional GAN (CGAN) model to generate poisoned data from local data. The architectures of the generator and discriminator of VagueGAN will be optimised in order to achieve superior poisoning results. Similar to the original CGAN, all data is generated through a multi-layer perceptron (MLP)-like network. Beyond the original CGAN, however, three enhancements are put in place for VagueGAN to make resulting poisoned data closer to legitimate ones in their distribution. First, VagueGAN utlizes a much deeper structure in the discriminator (e.g., 4 hidden layers). A deeper discriminator improves attack stealthiness by ensuring that the poisoned data better retains legitimate features even from a small local dataset. Second, VagueGAN also employs techniques such as batch normalization and LeakyReLU to more effectively capture data features. Third, VagueGAN adopts a large batch size for full dataset training to further obtain the overall features of local dataset. 

On the other hand, VagueGAN needs to generate noisy poisoned data for an effective attack, since data noise can easily harm the quality of training data and thus that of resulting trained model. First, besides the discriminator, VagueGAN also utlizes a much deeper structure in the generator (e.g., 4 hidden layers) and techniques such as
batch normalization and LeakyReLU. A deeper generator improves attack effectiveness by increasing the uncertainty of data generation and thus making the poisoned data vaguer and noisier. Second, we design a loss function of VagueGAN to break the commonly reached Nash equilibrium by restricting the discriminative ability of the discriminator. This can lower the generation performance of the generator and thus lead to generated data carrying a lot of poisonous noise.

Our loss function is defined as follows:
\begin{equation}
	\begin{split}
		\min_{G}\max_{D}V(G,D)=\mathbb{E}_{\bm{x}\sim p_{\text{d}}(\bm{x})}[\log{(1+\kappa)D(\bm{x})}] + \\
		\mathbb{E}_{\bm{z}\sim p_{\text{z}}(\bm{z})}[\log({1 -
			(1+\kappa)D(G(\bm{z}))})]
	\end{split}
\end{equation}
where $\bm{x}$ is a sample following the original data distribution $p_{\text{d}}(\bm{x})$; $\bm{z}$ is the sample obtained from a certain distribution $p_{\text{z}}(\bm{z})$; $G(\bm{z})$ is the fake sample generated by the generator $G$; $D(\bm{x})$ represents the probability that $\bm{x}$ is a real sample. On the one hand, the larger $D(\bm{x})$ (or $D(G(\bm{z})$) is, the more accurately the discriminator $D$ can identify original (or generated) samples. Hence, stronger $D$ is obtained to maximize $V$. On the other hand, if $G$ is stronger, the discriminator will make a misjudgment, and $D(G(\bm{z}))$ will become larger. Hence, stronger $G$ is obtained to minimize $V$. We use a suppression factor $\kappa \in (0,1) $ to limit the generation ability of the generator $G$. The larger $\kappa$ is, the more restrictive it is on the generation ability. Clearly, the value of $\kappa$ can be used to trade-off the effectiveness and stealthiness of VagueGAN.

\begin{figure}[h]
	\centering
	\includegraphics[width=\linewidth]{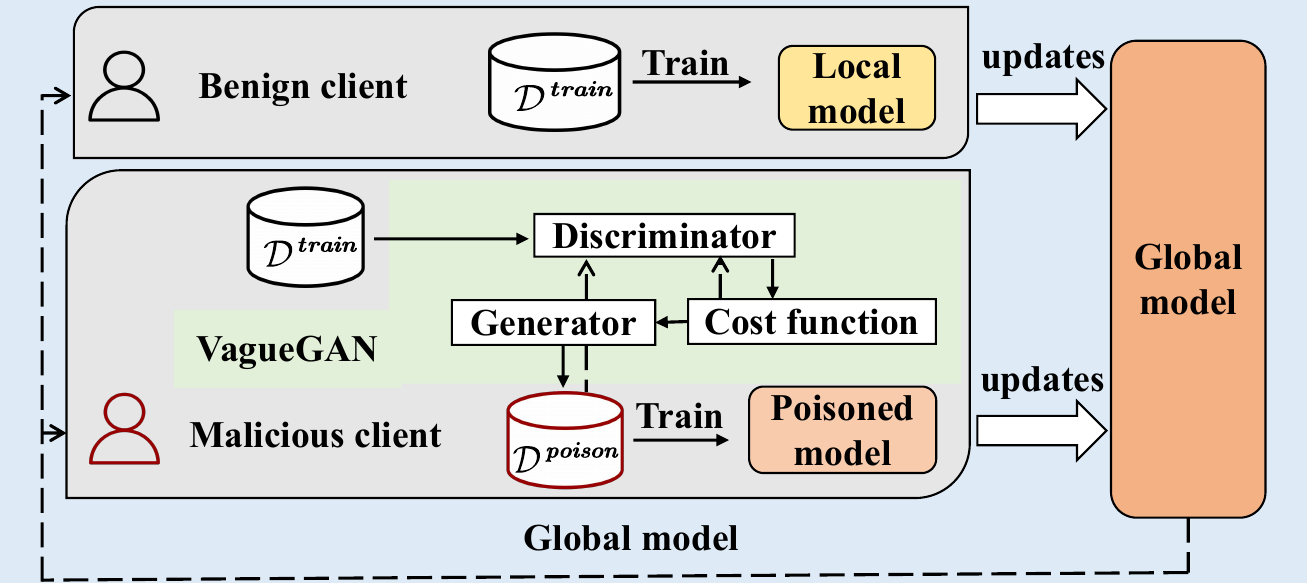}
	\caption{Use of VagueGAN for a data poisoning attack: A malicious client trains VagueGAN on its dataset to generate an equivalent amount of poisoned data. The poisoned dataset is then used for local training to obtain a poisoned model, which is subsequently uploaded to the server.}
\end{figure}

To support data poisoning attacks, on the one hand, the generation target of our VagueGAN is much different from that of traditional GANs for regular uses. As shown in Figure 1(b) and 1(g) where taking image data as an example, the sample distribution generated by a traditional GAN is largely fitted to the real sample distribution, but that generated by VagueGAN is just roughly close to the real sample distribution for a balanced trade-off between seemingly legitimate and noisy results. The samples generated by PoisonGAN (Figure 1(c)) with discriminator $D$ initialized by the global model~\cite{zhang2020poisongan} are also of relatively good quality because PoisonGAN can be considered as a data augmentation method instead of an actual attack method. Furthermore, compared with traditional GANs, our VagueGAN can be more computationally efficient only aiming to generate vague data of relatively low quality. Note that even if being under-trained for data poisoning, traditional GANs usually generate intermediate outputs with certain vagueness but without carrying noise, and thus cannot ensure successful attacks.

On the other hand, the vague data generated by VagueGAN is different from real data superimposed with noise (Figure 1(d), 1(e) and 1(f)). Adding noise directly to real data produces poisoned data whose attack quality is very difficult to control. Intuitively, better attack effectiveness can be achieved by making a noise level higher, but this can make the noise more noticeable and thus leading to worse attack stealthiness. This is because the manually added noise (e.g. Gaussian, Salt and Pepper (SAP)) has nothing related to the original data statistically. In contrast, the noise generation of VagueGAN is guided by the statistics of original data, so that the GAN-generated noise can more smoothly fit the original data. Our experiments will show that only VagueGAN can flexibly control the quality of vague data on demand to strike a balance between attack effectiveness and stealthiness. 

\subsection{VagueGAN Poisoning Attack}

Following the workflow of VagueGAN in Figure 2, the attacker first gains control over a client $c_j$ and deploys an untrained VagueGAN model. The compromised client then functions as a malicious participant in the FL process. Before federated training starts, the malicious client trains the VagueGAN model on its local dataset using the loss function defined in Eq. (1). The training process of VagueGAN is similar to those of traditional GANs, where the generator and discriminator undergo iterative optimization through an adversarial game. Next, the trained VagueGAN model generates poisoned dataset $\mathcal D^{\rm{poison}}_j$, which subsequently replaces the original dataset $\mathcal D_j$. Notably, the purpose of generating a poisoned dataset of the same size as the original dataset is just to ensure a fair comparison among different attack methods. That is, for every original data sample $(x, y)$, there exists exactly one corresponding poisoned sample $(x^p, y)$ in the poisoned dataset. Finally, client $c_j$ uses $\mathcal D^{\rm{poison}}_j$ to train and upload a poisoned local model $\bm{\theta}_{t,j}^{\rm{poison}}$ to the server. Unlike a traditional attack such as label flipping that only affects selected labels, our VagueGAN indirectly attacks a global model with all labels/classes involved. The steps for conducting a VagueGAN poisoning attack are summarized in Algorithm 1. 

\begin{figure}[!t]
	\renewcommand{\algorithmicrequire}{\textbf{Input:}}
	\renewcommand{\algorithmicensure}{\textbf{Output:}}
	\renewcommand{\algorithmicreturn}{\textbf{Initialization:}}
	\begin{algorithm}[H]
		\caption{VagueGAN poisoning attack algorithm}
		\begin{algorithmic}[1]
			\REQUIRE{Client $c_j$'s local training dataset $\mathcal D^{\rm{train}}_j$}
			\ENSURE{Poisoned local training dataset $\mathcal D^{\rm{poison}}_j$; local model parameters $\bm{\theta}_{t,j}^{\rm{poison}}$}
			\RETURN{Number of training epochs $E$ for VagueGAN; generator suppression factor $\kappa$; number of training rounds $T$ for FL}
			\STATE Deploy the VagueGAN model at a controlled client $c_j$ and receive an FL training task from the server.
			\STATE $//Poisoned$ $Data$ $Generation$
			\FOR{ $e = 1,2,...,E$} 
			\STATE Set training  batch  = $\mathcal D^{\rm{train}}_j$
			\STATE Train the VagueGAN model on $\mathcal D^{\rm{train}}_j$ according to (1) 
			\ENDFOR
			\STATE Get outputted $\mathcal D^{\rm{poison}}_j$
			\STATE $//FL$ $Training$ $Task$
			\FOR{$t = 1,2,...,T$}  
			\STATE Receive the global model $\bm{\tilde{\theta}}_{t}$ from the server 
			\STATE Train the local model $\bm{\theta}_{t,j}$ on $\bm{\tilde{\theta}}_{t}$ and $\mathcal D^{\rm{poison}}_j$
			\STATE Get outputted $\bm{\theta}^{\rm{poison}}_{t+1,j}$ and upload it to the server 
			\ENDFOR
		\end{algorithmic}
	\end{algorithm}
\end{figure}
\subsection{Theoretical Analysis}

This subsection theoretically reveals the characteristics of poisoned data generated by VagueGAN. For the expectation form of the loss function in Eq. (1), we rewrite it in integral notation as follows:
\begin{equation}
	\begin{split}
		\min_{G}\max_{D}V ( G , D ) = \int _ { \boldsymbol{x} } {p_d}  ( \boldsymbol{x} ) \log (1+\kappa)D ( \boldsymbol{x} ) d \boldsymbol{x} + \\
		\int _ { \boldsymbol{z} } {p_{\text{z}}} ( \boldsymbol{z} ) \log ( 1 - (1+\kappa)D ( G ( \boldsymbol{z} ) ) ) d \boldsymbol{z}.
	\end{split}
\end{equation}
The data distribution learned by the generator $G$ is represented as $p_g(\boldsymbol{x})$. According to the Law of the Unconscious Statistician (LOTUS) theorem, by replacing $\boldsymbol{z}$ using its mapping to $\boldsymbol{x}$ and then transforming it into an integral form, we can obtain an alternative expression for Eq. (4):
\begin{equation}
	\begin{split}
		\min_{G}\max_{D}V ( G , D ) = \int _ { \boldsymbol{x} } [{p _  d} ( \boldsymbol{x} ) \log (1+\kappa)D ( \boldsymbol{x} ) + \\
		p _ { g } ( \boldsymbol{x} ) \log ( 1 - (1+\kappa)D ( \boldsymbol{x} ) ) ]d \boldsymbol{x}.
	\end{split}
\end{equation}

For a fixed generator $G$, our objective is to solve for $\max_{D}V ( G , D )$, which entails finding the optimal discriminator $D^*$ that maximizes the right-hand side of the equation. 
\begin{equation}
	\begin{split}
		\max_{D}V ( G , D ) = \int _ { \boldsymbol{x} } [{p _  d} ( \boldsymbol{x} ) \log (1+\kappa)D ( \boldsymbol{x} ) + \\
		p _ { g } ( \boldsymbol{x} ) \log ( 1 - (1+\kappa)D ( \boldsymbol{x} ) ) ]d \boldsymbol{x}.
	\end{split}
\end{equation}
In Eq. (4), $p_d(\boldsymbol{x})$ refers to the probability of sampling a sample $\boldsymbol{x}$ from the distribution $p_{d}$, $p_g(\boldsymbol{x})$ refers to the probability of sampling a sample from the generator $G$, and $D(\boldsymbol{x})$ represents a mapping of $x\in \mathbb{R}$. The maximization of the integral in Eq. (4) can be reformulated as finding the maximum value of the integrand, through determining the optimal discriminator $D$. Consequently, during this process, the probabilities associated with the real data distribution $p_d(\boldsymbol{x})$ and the generated data distribution $p_g(\boldsymbol{x})$ can be treated as constants. Taking the partial derivatives of both sides of the equation with respect to $D$, we obtain:
\begin{equation}
	\begin{split}
		\frac{\partial}{\partial D}(\max_{D} V(G,D))= \int _ { \boldsymbol{x} } \frac{\partial}{\partial D}[ p_{d}(\boldsymbol{x}) \log ((1+\kappa)D(\boldsymbol{x}))\\
		+ p_{g}(\boldsymbol{x})\log (1-(1+\kappa)D(\boldsymbol{x}))] d\boldsymbol{x} \\
		= \int _ { \boldsymbol{x} }\left[ p_d(\boldsymbol{x}) \frac { 1 } { D(\boldsymbol{x}) } + p_g(\boldsymbol{x}) \frac { - (1+\kappa) } { 1 - (1+\kappa) D(\boldsymbol{x}) } \right] d \boldsymbol{x}.
	\end{split}
\end{equation}
By taking further derivatives of Eq. (5), it can be determined that Eq. (4) is a convex function. Therefore, Eq. (4) attains a unique maximum value at the point where its first derivative is equal to zero. At this point: 
\begin{equation}
	D^*\left( \boldsymbol{x} \right) = \frac{p_d\left( \boldsymbol{x} \right)}{(1+\kappa)(p_d\left( \boldsymbol{x} \right) +p_g\left( \boldsymbol{x} \right))}.
\end{equation}
When $D^*(\boldsymbol{x})=0.5$, the specially designed discriminator in VagueGAN is unable to distinguish between real samples and fake samples generated by the generator. At this point, the generator of VagueGAN reaches its generation bottleneck. Consequently, one can conclude that:
\begin{equation}
	p_g\left( \boldsymbol{x} \right) =\frac{\left( 1-\kappa \right)}{(1+\kappa)} p_d\left( \boldsymbol{x} \right).
\end{equation}

Eq. (7) reveals that VagueGAN has a distinct generation objective compared to traditional GANs. When VagueGAN achieves optimal performance, the distribution of the generated poisoned data is related to the original data distribution, but not identical. This distinction arises from the unique suppression factor $\kappa$ in VagueGAN. By adjusting the suppression factor $\kappa$, VagueGAN can control the level of conformity between the generated poisoned data and the real data. Within the feasible range of $\kappa$ ($\kappa \in (0,1)$), a higher value of $\kappa$ leads to a reduced correlation between the poisoned data and the real data, while a lower value of $\kappa$ yields a higher correlation. When $\kappa$ equals 0, VagueGAN is equivalent to a regular GAN.

\begin{figure}[h]
	\centering
	\includegraphics[width=\linewidth]{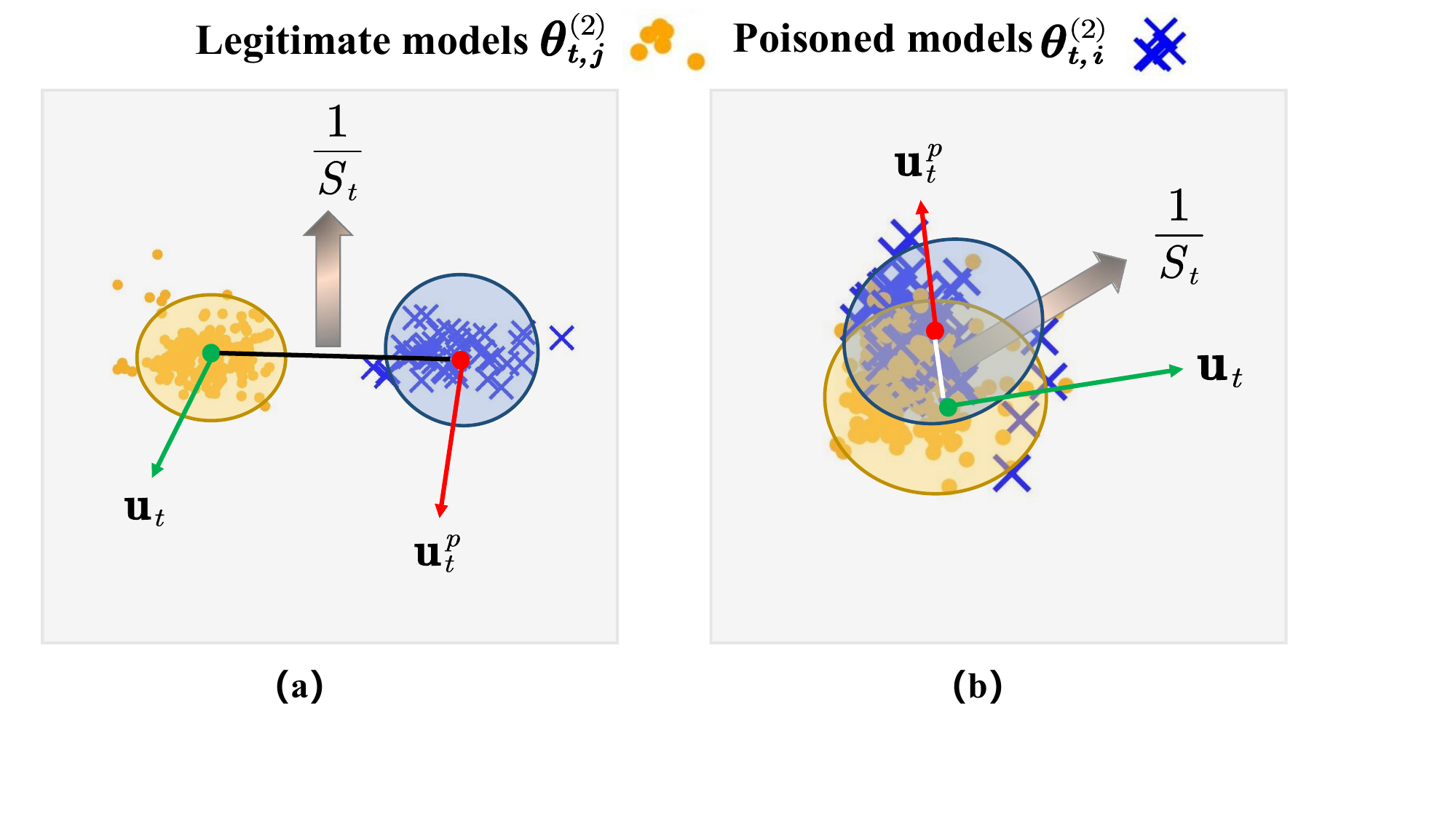}
	\caption{Different levels of attack stealthiness: (a) Lower attack stealthiness for larger inter-cluster distance; (b) Higher attack stealthiness for smaller inter-cluster distance. }
\end{figure}

\begin{figure}[h]
	\centering
	\includegraphics[width=\linewidth]{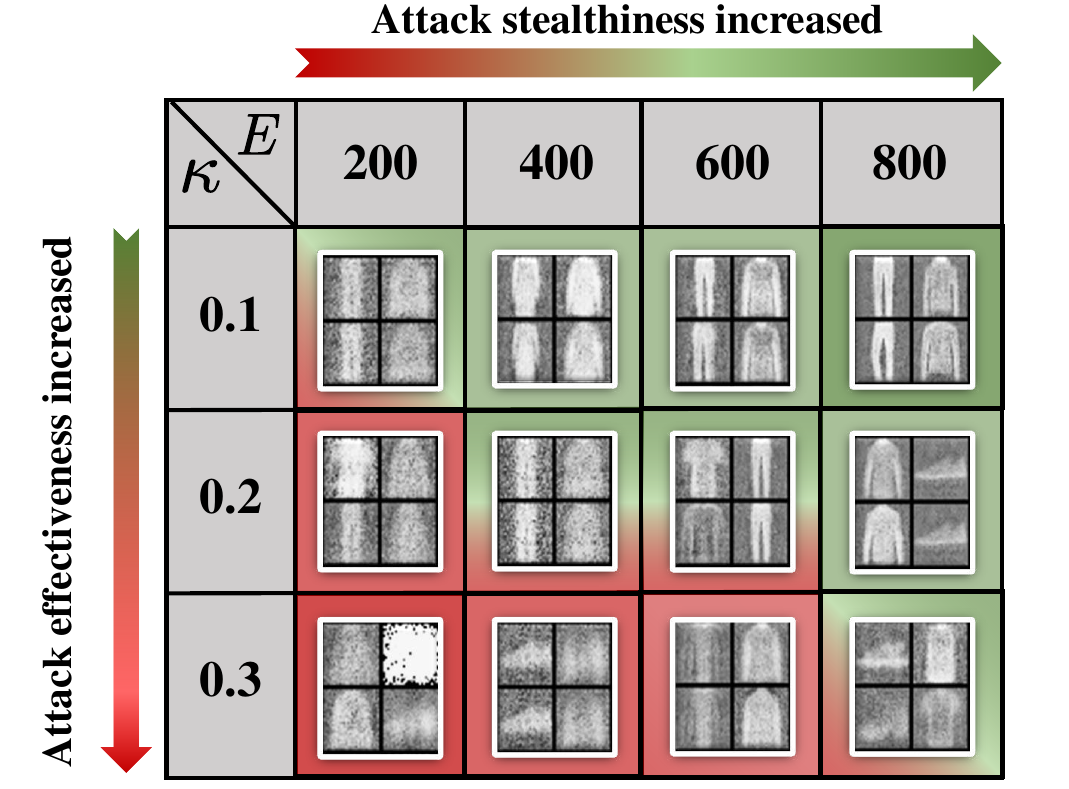}
	\caption{Trade-off between attack effectiveness and stealthiness: Higher effectiveness but lower stealthiness for the lower left region, while lower effectiveness but higher stealthiness for the upper right region. }
\end{figure}

\subsection{Attack Effectiveness-Stealthiness Trade-off}

There is usually a trade-off between attack effectiveness and stealthiness. Noisier data generated by VagueGAN can be more effective in degrading FL performance, but may bring greater statistical changes and thus may be easier to detect. Therefore, the attacker needs to make appropriate settings for VagueGAN to fine-tune the quality of vague data and achieve a balanced trade-off between attack effectiveness and stealthiness. Here we give some guidelines for taking full advantage of VagueGAN for the balanced trade-off. 

The effectiveness $A_t$ of an attack in round $t$ of FL can be characterized by the drop in the accuracy of the global model from $a_t$ to $a^p_t$ due to the presence of the attack:
\begin{equation}
	A_t= a_t- a_t^{p}.
\end{equation}
For a multi-class classification task, we employ the one-vs-rest strategy to compute the overall accuracy metric $ a_t $:
\begin{equation}
a = \frac{\sum_{i=1}^{l} (TP_{t,i} + TN_{t,i})}{\sum_{i=1}^{l} (TP_{t,i} + TN_{t,i}+ FP_{t,i} + FN_{t,i})}
\end{equation}
where $ l $ denotes the total number of classes. For each class $ i $, if we treat it as the positive and all the other classes as the negative, we can easily compute the numbers of true positives ($ TP_i $), true negatives ($ TN_i $), false positives ($ FP_i $), and false negatives ($ FN_i $), respectively, and finally the overall accuracy.

The stealthiness $S_t$ of an attack in round $t$ of FL can be characterized by the statistical distance between the clusters of legitimate and poisoned models. Since every local model $\bm{\theta}_{t,i}$ is very high-dimensional, we first use PCA to downscale $\bm{\theta}_{t,i}$ to 2-dimensional $\bm{\theta}_{t,i}^{(2)}$.
\begin{equation}
	\bm{\theta}_{t,i}^{(2)}=\mathrm {PCA} (\bm{\theta}_{t,i}).
\end{equation}
Then, we separate all legitimate local models and poisoned local models into two clusters and find the centroids $\mathbf u_t$ and $\mathbf u_t^p$ of these two clusters, respectively. 
\begin{equation}
	\mathbf u_t=\frac{\sum_{i=1}^{N-M}\bm{\theta}_{t,i}^{(2)}}{N-M}
\end{equation}
\begin{equation}
	\mathbf u^p_t=\frac{\sum_{j=1}^{M}\bm{\theta}_{t,j}^{(2)}}{M}
\end{equation}
where $N$ is the number of all clients and $M$ is the number of malicious clients, so $N-M$ is the number of benign clients. We use the reciprocal of the Euclidean distance $\mathbf{E}$ between $\mathbf u_t$ and $\mathbf u_t^p$ to measure $S_t$:
\begin{equation}
	S_t=\frac{1}{\mathbf{E}(\mathbf u_t, \mathbf u^p_t)}=\frac{1}{\lVert \mathbf u_t-\mathbf u^p_t \rVert}.
\end{equation}
Specifically, this study employs the centroid linkage definition under the Euclidean distance metric. Note that it also supports other inter-cluster distance definitions such as single linkage, complete linkage, or average linkage, or other distance metrics such as cosine similarity. Figure 3 illustrates two cases with different levels of attack stealthiness. The performance of $S_t$ is lower (or higher) as in Figure 3(a) (or 3(b)) when the statistical distance between legitimate and poisoned models is larger (or smaller). 

To study the trade-off between attack effectiveness $A_t$ and stealthiness $S_t$, we have analyzed the experimental outputs of VagueGAN, as in Figure 4. Clearly, the trade-off is closely related to the quality of poisoned data generated by VagueGAN. The more VagueGAN learns the characteristics of legitimate data, the higher stealthiness but lower effectiveness  the resulting attack can achieve, and vice versa. In Figure 4, there are three general attack outcomes given by VagueGAN: low effectiveness/high stealthiness (upper right region), high effectiveness/low stealthiness (lower left region), and balanced effectiveness/stealthiness (central region). Due to the mutually exclusive nature of effectiveness and stealthiness, achieving both high effectiveness and high stealthiness by an attack is relatively unfeasible. Hence, the balanced outcome can be a rational goal for the attacker.  

In follow-up experiments, it has been found that there are two important factors affecting the quality of poisoned data: VagueGAN training epochs $E$ and generator suppression factor $\kappa$. A balanced trade-off between attack effectiveness $A_t$ and stealthiness $S_t$ can be achieved by setting the values of $E$ and $\kappa$ appropriately. $E$ is set to make VagueGAN stay in a semi-fitting state, and $\kappa$ is adjusted to limit the generation ability of the generator of VagueGAN. For example, poisoned data of balanced quality around the middle area in Figure 4 can be generated by experimentally setting $E\in[400,600]$ and $\kappa=0.2$. This configuration aims to achieve the ideal state of median effectiveness/median stealthiness. More details can be found in subsection VI-C. 
\begin{figure}[h]
	\centering
	\includegraphics[width=\linewidth]{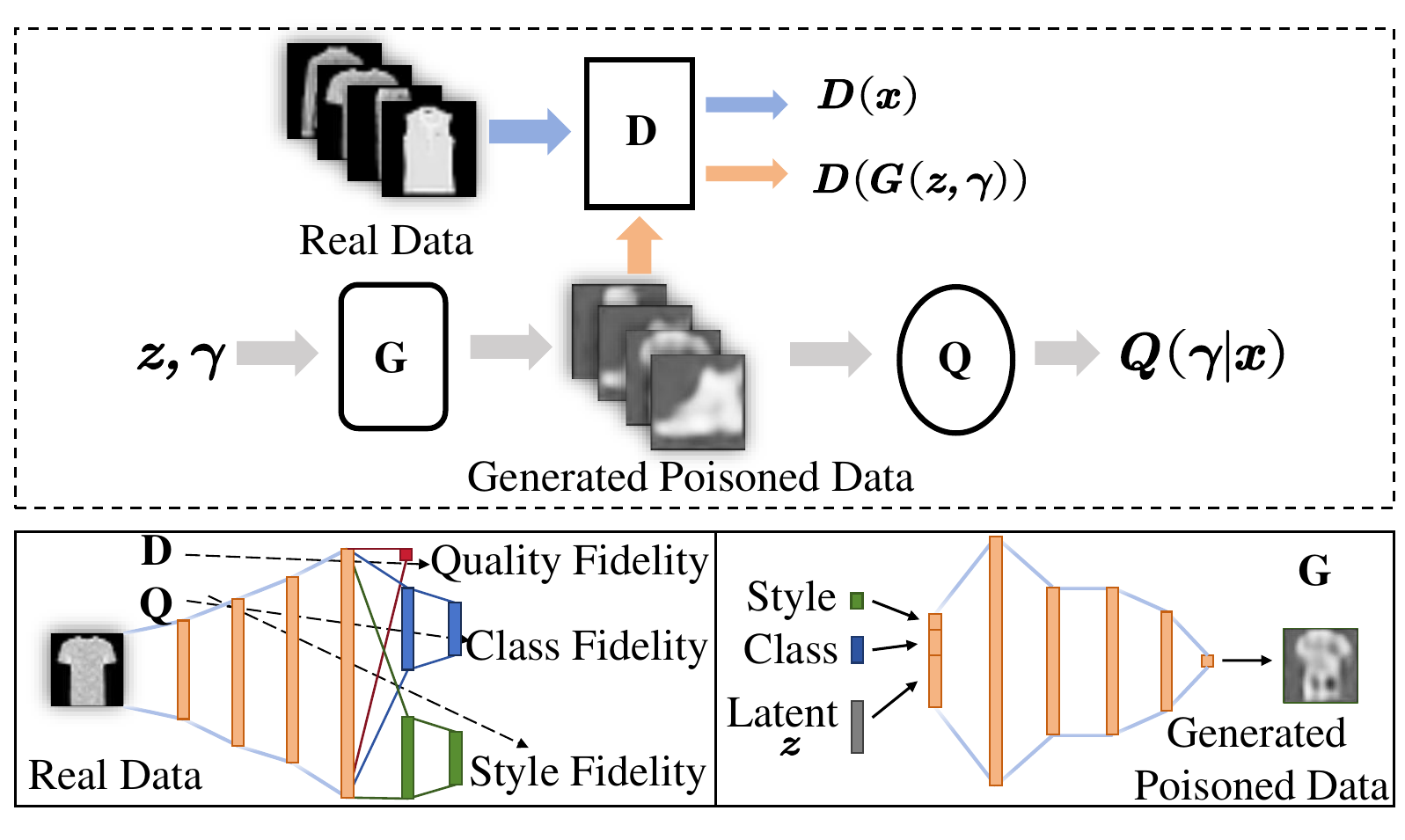}
	\caption{Architecture of unsupervised VagueGAN: An extra classifier network $Q$ measures the validity of auxiliary vector. }
\end{figure}
\begin{figure*}[h]
	\centering
	\includegraphics[width=\linewidth]{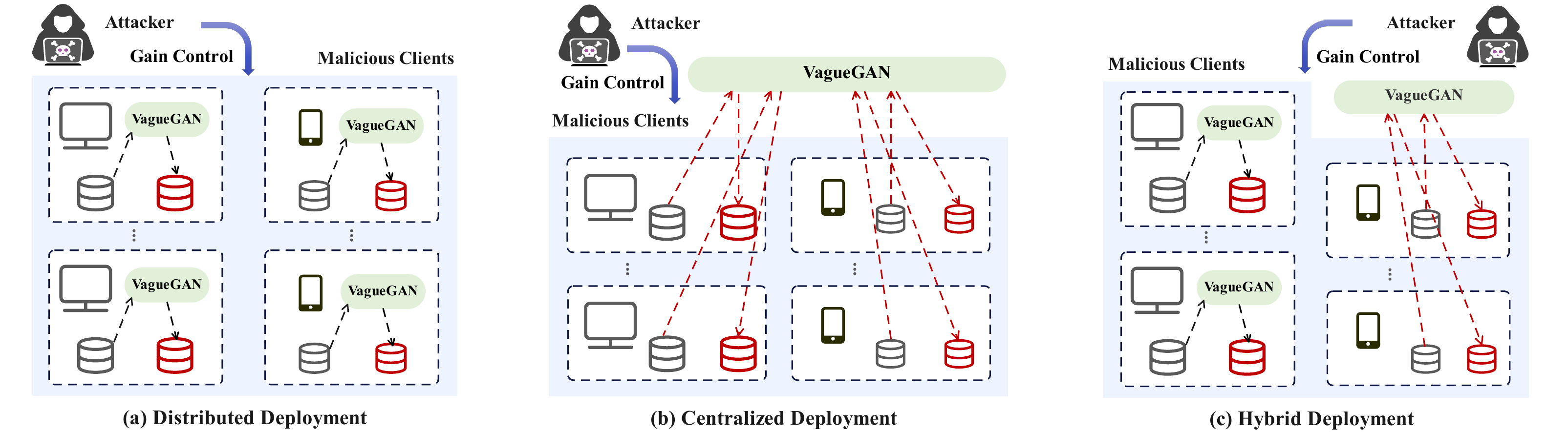}
	\caption{Three ways to deploy VagueGAN: The attacker can flexibly deploy VagueGAN on demand. }
	\vspace{-10pt}
\end{figure*}

\subsection{Unsupervised Variant of VagueGAN}

%Given that clients in FL may not always possible to prepare accurate labels of training samples beforehand. Sometimes, it is even difficult for us to accurately describe the labels of extremely complex data. Inspired by InfoGAN, VagueGAN also provides unsupervised generation of poisoned data. Similar to supervised generation, unsupervised VagueGAN also replaces the original distribution of data with conditional distribution (with auxiliary information as conditions). The main difference is that unsupervised VagueGAN does not need to feed label and attribute information into the discriminator network; instead, it uses another classifier, $Q$, to measure how auxiliary features are learned. The objective function of unsupervised VagueGAN is as follows:

Sometimes the attacker may not have full access to the label information of a local dataset, making the generation of poisoned data even more challenging. Inspired by InfoGAN, we further propose an unsupervised variant of VagueGAN for broader applicability. As shown in Figure 5, unsupervised VagueGAN could not only learn to generate realistic samples but also learn auxiliary features (class fidelity and style fidelity). Similar to supervised generation, unsupervised VagueGAN also replaces the original distribution of data with a conditional distribution. The main difference lies in that unsupervised VagueGAN does not require feeding label and attribute information into the discriminator network. Instead, it utilizes another classifier $Q$ to learn auxiliary features. 

As shown in Figure 5, both the discriminator and generator networks of unsupervised VagueGAN consist of 4 hidden layers. Unsupervised VagueGAN uses an additional classifier network $Q$ to measure the validity of the auxiliary vector. This additional classifier shares most of its weights (the first 4 hidden layers) with the discriminator.

The objective of unsupervised VagueGAN is defined as follows, with an addition $-\lambda L\left( G,Q \right)$, to Eq. (1):
\begin{equation}
	\begin{split}
		\min_{G,Q}&\max_{D}V(G,D,Q)=\mathbb{E}_{\bm{x}\sim p_{\text{d}}(\bm{x})}[\log{(1+\kappa)D(\bm{x})}] + \\
		&\mathbb{E}_{\bm{z}\sim p_{\text{z}}(\bm{z}),\gamma \sim p\left( \gamma \right)}[\log({1 -
			(1+\kappa)D(G(\bm{z,\gamma}))})] \\ &-\lambda L\left( G,Q \right) 
	\end{split}
\end{equation}
where $\lambda$ is hyperparameter, $x\sim G\left( z,\gamma \right)$ is the generated poisoned data, and $\gamma$ represents auxiliary information, which includes class fidelity and style fidelity generated from classifier $Q$. $p\left( \gamma \right)$ describes the actual distribution of $\gamma$, which is rather hard to obtain. Therefore, VagueGAN uses the posterior distribution, $Q\left( \gamma |x \right)$, to estimate $p\left( \gamma \right)$, and this process is done with a neural network classifier. $L\left( G,Q \right)$ is an approximation of mutual information (lower bound of mutual information), $I(\gamma ;G\left( z,\gamma \right)) $, between the auxiliary vector and generated poisoned sample: 
\begin{equation}
	\begin{split}
		L\left( G,Q \right) =&\mathbb{E}_{\gamma \sim p\left( \gamma \right) ,x\sim G\left( z,\gamma \right)}\left[ \log Q\left( \gamma |x \right) \right] +H\left( \gamma \right) \\
		=&\mathbb{E}_{x\sim p_G\left( x|z,\gamma \right)}E_{\gamma \sim p\left( \gamma |x \right)}\log Q\left( \gamma |x \right) +H\left( \gamma \right) \\
		\leq& I\left( \gamma;G\left( z,\gamma \right) \right) . 
	\end{split}
\end{equation}
Mutual information, $I\left( \gamma;G\left( z,\gamma \right) \right) $, describes how much VagueGAN knows about random variable $\gamma$ based on knowledge of $G\left( z,\gamma \right)$, i.e. $I\left( \gamma;G\left( z,\gamma \right) \right) =H\left( \gamma \right) -H\left( \gamma|G\left( z,\gamma \right) \right)$, in which $H(\gamma)$ is conditional entropy. It can also be described by the Kullback-Leibler divergence, the information loss when using marginal distributions to approximate the joint distribution of $\gamma$ and $G\left( z,\gamma \right)$. %Thanks to the unique loss function of VagueGAN, the effectiveness and stealthiness of supervised and unsupervised VagueGAN attacks are quite similar. However, the unsupervised VagueGAN can generate poisoned data in scenarios with limited resources, such as noisy or missing client labels.
%Equation (9) represents the upper limit on the quality of poisoned data generated by the generator $G$ in VagueGAN. It is evident that when VagueGAN achieves its optimal performance, the distribution of poisoned data generated by the generator is correlated with the original data distribution, albeit not entirely identical. By adjusting the suppression factor $\kappa$, we can regulate the quality of poisoned data generated by VagueGAN. Within the feasible range of $\kappa$($\kappa \in (0,1) $), a larger value of $\kappa$ leads to a decreased correlation between the poisoned data and the real data, whereas a smaller value of $\kappa$ results in a closer approximation.

Thanks to the unique loss function, the effectiveness and stealthiness of supervised and unsupervised VagueGAN attacks are comparable. Detailed evaluation results can be found in subsection VI-D.
\subsection{Deployment Advantages}

%Furthermore, the VagueGAN model does not necessitate strict Nash equilibrium attainment.   Consequently, implementing a data poisoning attack through VagueGAN is straightforward and feasible.     This poisoned data is utilised throughout the FL training process.    Due to variations in application contexts,      VagueGAN and the original GAN manifest notable differences across multiple aspects.
VagueGAN can be readily deployable in practice. First, VagueGAN is relatively easy to train. VagueGAN introduces a unique loss function that prevents from generating high-quality data unnecessary for attack purpose. In addition, VagueGAN does not have to wait for the convergence to a Nash equilibrium to obtain relatively low-quality but damaging poisoned data, leading to substantially reduced training efforts and costs compared to regular GANs. The adoption of full dataset training also demands fewer epochs to generate expected poisoned data. The limitations inherent to a regular GAN, such as non-convergence and unstable training, do not hinder the use of VagueGAN. In comparison, the original GAN is unable to generate data that meets the requirements for poisoning attacks. Even its intermediate outputs during the generation process can only achieve a degree of "vagueness" without carrying "noise". Additionally, even with simpler models, achieving its generation goal is associated with high training costs. Second, VagueGAN only incurs one-time cost. It requires to build up a loss function from available datasets only once, enabling the generation of an arbitrary amount of poisoned data thereafter. Third, VagueGAN can be flexibly deployed depending on the attacker's resource availability. In general, there are three ways to put VagueGAN into use, including distributed deployment, centralized deployment, and hybrid deployment, so it is not hard for the attacker to find a way to deploy VagueGAN.  

%VagueGAN introduces a unique loss function that constrains the upper bound of data generation, which does not necessitate strict Nash equilibrium attainment, leading to substantially reduced training efforts and costs compared regular GANs. The limitations inherent to a regular GAN, such as non-convergence and unstable training, do not hinder the use of VagueGAN. Because VagueGAN is engineered to generate vague and uncontrollable poisoned data. VagueGAN requires building up a loss function from available datasets only once, enabling the generation of an arbitrary amount of poisoned data. In order to empower attackers to achieve optimal attack effectiveness using VagueGAN, tailored to the main task and controlled client conditions, we have formulated three following distinct attack strategies:

\textbf{Distributed Deployment:} As shown in Figure 6(a), if the attacker has gained the control of one or more clients with rich on-device resources, a VagueGAN model can be individually trained on each of the malicious clients, giving diversified poisoned datasets. When we use the term "VagueGAN" in this paper, we will mean VagueGAN implementing a distributed deployment, unless we state otherwise.
%each controlled client autonomously trains VagueGAN in isolation, making this the most optimal attack strategy when the controlled clients are resource-ample. The poisoned data generated by each controlled client remains independent, with no mutual influence. Throughout this process, there is no introduction of abnormal communication beyond the confines of the FL process.
%Notably, the poisoned dataset maintains certain personalized features inherent in each original dataset.

\textbf{Centralized Deployment:} As shown in Figure 6(b), if the attacker has access to a resource-rich platform reachable from the clients, one or more VagueGAN models can be centrally trained after collecting local data from the malicious clients. Then either unified or diversified poisoned datasets can be distributed to the malicious clients. 
% Implementing VagueGAN on the attacker's device eliminates the need for controlled clients to undergo VagueGAN training in isolation.

\textbf{Hybrid Deployment:} As shown in Figure 6(c), if some of the malicious clients have rich on-device resources but others do not, a mixture of locally trained and centrally trained VagueGAN models can be applied. 
%training a VagueGAN utilizes resources comparable to a single FL task. Generally, 

%In reality, the expense of building up a loss function in VagueGAN is significantly low, and it is adequate to train a functional VagueGAN model as long as a controlled client can complete federated training. In most instances, an isolation strategy is feasible. Two supplementary strategies aim to make VagueGAN more adaptable and pragmatic in idiosyncrasies and real-world situations. When we use the term "VagueGAN", we will mean VagueGAN implementing an isolation strategy, unless we state otherwise.

%.Given the idiosyncrasies real-world stiuations and uncertain external factors, VagueGAN implements two supplementary strategies to augment its adaptability and pragmatically.

\section{Model Consistency-Based Defense}

Thanks to its inherent generative capability, our VagueGAN can generate seemingly legitimate poisoned data and thus can bypass existing defense methods. In this section, we analyze the characteristics of GAN outputs and suggest a model consistency-based countermeasure to stealthier GAN-based data poisoning attacks. Our defense method is effective when the same VagueGAN model is consistently applied by the attacker, though the idea of VagueGAN still has great potential to deliver undetectable attacks if implemented in varying settings. Moreover, our defense method also works well against mainstream data poisoning attacks.  

\subsection{Local Model Analysis}

To find a clue for the identification of malicious clients given that the server does not have access to any local data, we analyze the statistical patterns of local models successively uploaded by a certain client within a certain time span. 

Due to the high-dimensional characteristics of local models, it would be difficult to review them directly and the review process would impose a huge computational load on the server. To better analyze the local models, we use PCA to reduce the high-dimensional models to 2-dimensional ones to extract the most useful information. Typical distributions of legitimate and poisoned models after PCA are compared in Figure 7, and the corresponding experimental details can be found in section VI.

\begin{figure}[h]
	\centering
	\includegraphics[width=\linewidth]{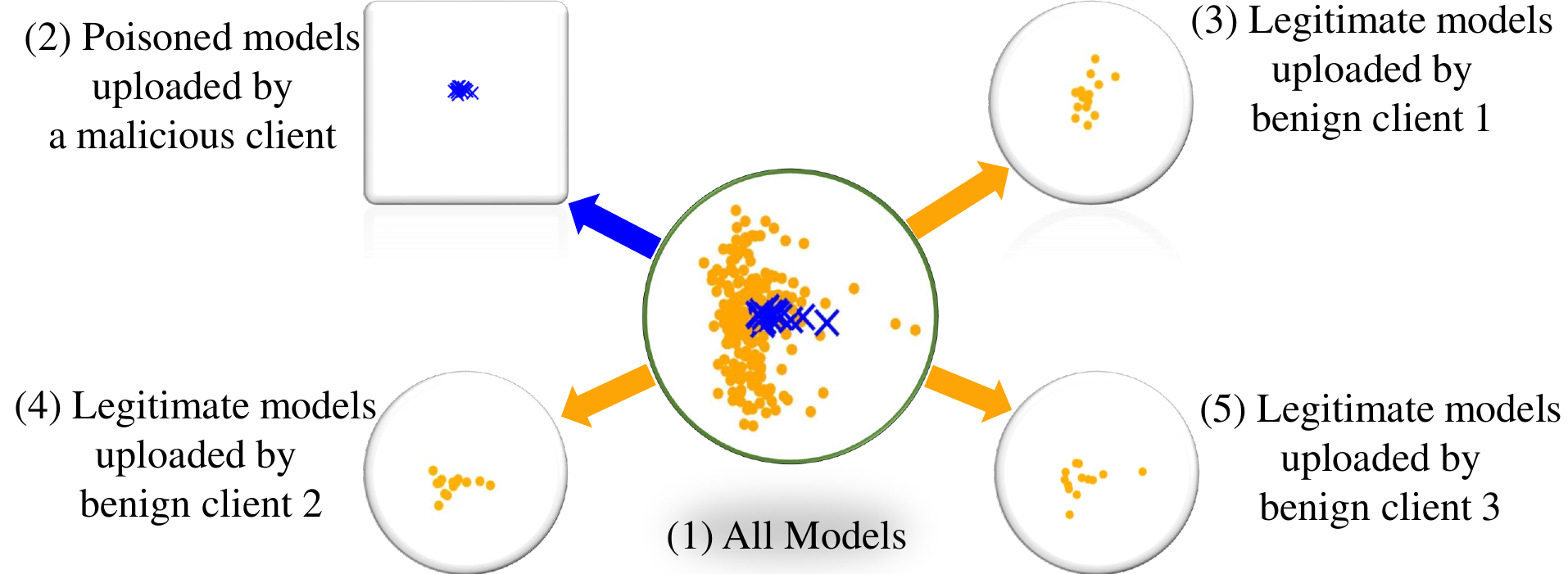}
	\caption{Legitimate and poisoned models after PCA: Higher model consistency can be observed for poisoned models compared with legitimate ones. }
\end{figure}  

It can be seen that the distribution of poisoned models given by the VagueGAN model is close to those of legitimate models except for their distribution density or consistency levels. The poisoned models uploaded by the same malicious client are more densely distributed within a smaller region. This is mainly caused by the consistency of GAN outputs, as GAN-generated data usually have a certain degree of data feature similarity. Malicious clients keep using similar GAN-poisoned data train poisoned models, and the consistency of GAN outputs further leads to the consistency of the corresponding poisoned models. Inspired by this observation, we propose to identify malicious clients by examining the consistency level of local models successively uploaded by every client. 

\subsection{MCD Design}
%Table IV provides the notation utilized in MCD in this paper.

In order to address the stealthier data poisoning attacks, we propose a new defense method named Model Consistency-Based Defense (MCD), which is performed by the server in four steps. The specific process is as follows:
%Step 1 model preprocessing: MCD preprocesses the local models into two metrics to prepare them for subsequent analysis and computation. Step 2 baseline value calculation: MCD calculates the baseline values of the two metrics for anomaly detection. The baseline values reflect the level of consistency among the local models. Step 3 client abnormality calculation: Using the baseline values, MCD calculates every trusted client abnormality. The abnormality measures the degree of deviation of each client from the others. Step 4 malicious client identification: MCD marks clients with high abnormality as malicious ones. Table IV provides the notation utilized in MCD in this paper. The specific process is as follows:

\paragraph{\textbf{Step 1 (Model Processing)}}MCD maintains a trusted set of clients, denoted as $C^{\rm{tr}}$. Here, $C^{\rm{tr}}$ refers to a collection of clients that are temporarily considered to be benign in the system. During the initialization phase, $\mathcal C^{\rm{tr}}$ is set to $\mathcal C$. MCD runs periodically for every defense period of $T'$ FL rounds for local model analysis. In each defense period, MCD stores a set $\mathcal Q_{f,i}$ of low-dimensional local models uploaded by every trusted client $c_i \in C^{\rm{tr}}$ to the server in $T'$ rounds, $f=0,1,2,3,...$ stands for the $f$th defense period, thus $f=\lfloor \dfrac{t}{T'} \rfloor $. Here we take 2-dimensional models after dimensionality reduction by PCA as an example, but MCD is not limited to it. In our FL system, the server randomly selects $K$ out of $N$ clients per round for model aggregation, so $\bm{\theta}_{t,i}^{(2)} \in \mathcal Q_{f,i}$ is set to $NULL$ if client $c_i$ is not selected in round $t$. MCD identifies malicious clients and removes them from $\mathcal C^{\rm{tr}}$ in each defense. 

\begin{figure}[h]
	\centering
	\includegraphics[width=\linewidth]{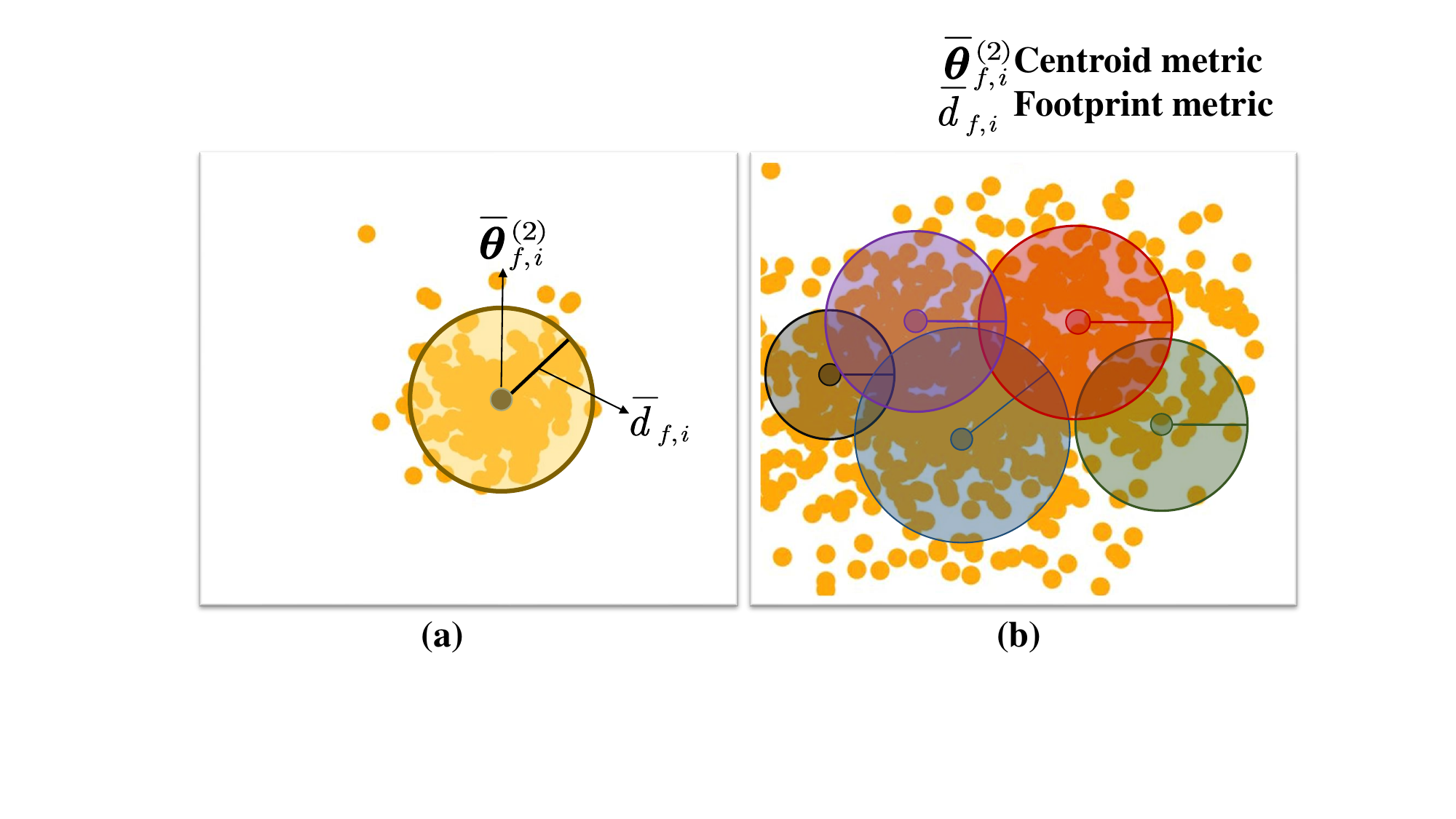}
	\caption{Two metrics for model consistency evaluation: A centroid metric $\boldsymbol{\bar{\theta}}_{f,i}^{(2)}$ and a footprint metric $\bar{d}_{f,i}$.}
\end{figure}

Afterward, MCD utilizes two crucial metrics calculated from $\mathcal Q_{f,i}$ of each trusted client $c_i$ to evaluate a model consistency level. First, for each trusted client $c_i$, MCD calculates a centroid metric :
\begin{equation}
	\boldsymbol{\bar{\theta}}_{f,i}^{(2)}=\frac{\sum_{\boldsymbol{\theta}_{t,i}^{(2)}\in \mathcal{Q}_{f,i} }{\boldsymbol{\theta }_{t,i}^{(2)}}}{\left| \mathcal{Q}_{f,i} \right|}
\end{equation}
where $\boldsymbol{\bar{\theta}}_{f,i}^{(2)}$ represents the center of all models uploaded by a single client $c_i$ (shown in Figure 8(a)). Anomalies with the centroid metric can be used to detect most of the existing data poisoning attacks, such as label flipping attacks, PoisonGAN, noise superimposition attack, etc, as validated in subsection VI-D. Second, MCD calculates a footprint metric 
\begin{equation}
	\bar{d}_{f,i}=\frac{\sum_{\boldsymbol{\theta }_{t,i}^{\left( 2 \right)}\in \mathcal{Q}_{f,i}}{\lVert \boldsymbol{\theta }_{t,i}^{(2)}-\boldsymbol{\bar{\theta}}_{f,i}^{(2)} \rVert}}{\left| \mathcal{Q}_{f,i} \right|}
\end{equation}
where $\bar{d}_{f,i}$ is the mean of the Euclidean distances, i.e., variances, of all models from the centroid $\boldsymbol{\bar{\theta}}_{f,i}^{(2)}$. A circular region defined by $\boldsymbol{\bar{\theta}}_{f,i}^{(2)}$ as the center and $\bar{d}_{f,i}$ as the radius represents a simplified footprint of the client $c_i$'s model distribution (shown in Figure 8(a)). Anomalies with the footprint metric can be used to detect GAN-based data poisoning attacks when the same GAN model like VagueGAN is applied consistently. Each pair of centroid and footprint metrics is extracted from a set of high-dimensional local models, and MCD can cost-effectively work with such extracted metrics instead of the original models. This  greatly saves the storage space of the server, and also keeps the computational effort of MCD low to identify malicious clients. 

Thus far, MCD pre-processes each trusted client $c_i$'s local models into a pair of two key metrics $(\boldsymbol{\bar{\theta}}_{f,i}^{(2)}, \bar{d}_{f,i})$. Figure 8(b) presents an example of five clients' metric pairs.
%, and the set composed of the metric pairs of all trusted clients is denoted as $\mathcal{U}$.

%As an example, Figure 7(a) shows the definitions of $\boldsymbol{\bar{\theta}}_{t,i}^{(2)}$ and $\bar{d}_{t,i}$, and Figure 7(b) shows accordingly simplified model distributions for five clients.  

%After calculating $(\boldsymbol{\bar{\theta}}_{t,i}^{(2)}, \bar{d}_{t,i})$ for all client models, MCD uses the model metrics set $\mathcal U_i=(\boldsymbol{\bar{\theta}}_{t,i}^{(2)}, \bar{d}_{t,i})$ to represent the model information stored in $\mathcal Q_i$. Then server no longer stores a large number of high-dimensional local models, 

\begin{algorithm}[tb]   
	\caption{Client abnormality evaluation algorithm} 
	\label{alg:algorithm} 
	\textbf{Input}: Trusted client set $\mathcal C^{\rm{tr}}$; baseline client set $\mathcal C^{\rm{base}}$\\
	\textbf{Output}: All trusted clients' abnormality set $\mathcal H_f$\\
	\textbf{Initialization}: Weights of two anomalous terms $\lambda _1$, $\lambda _2$\\
	\begin{algorithmic}[1]
		\STATE $//Calculate$ $Baseline$ $Values$
		\FOR{ $ c_i\in\mathcal C^{\rm{base}}$} 
		\STATE Calculate the base value $\boldsymbol{{\theta}}^{\rm{base}}_f$ according to (20)
		\STATE Calculate the base value ${d}_{f}^{\rm{base}}$ according to (21) 
		\ENDFOR
		\STATE $//Calculate$ $Abnormal$ $Degree$
		\FOR{$c_i\in\mathcal C^{\rm{tr}}$}  
		\STATE Calculate the client abnormality $h_{f,i}$ with $\lambda _1$, $\lambda _2$ according to (21) 
		\STATE Get outputted trusted client $c_i$'s abnormality $h_{f,i}$ 
		\ENDFOR
		\STATE Output all trusted clients' abnormality set $\mathcal H_f$
	\end{algorithmic} 
\end{algorithm}

\begin{algorithm}[tb]   
	\caption{MCD defense algorithm} 
	\label{alg:algorithm} 
	\textbf{Input}: All clients' model set $\mathcal Q_f=\{\mathcal Q_{f,1},\mathcal Q_{f,2},...Q_{f,i},...,Q_{f,N}\}$\\
	\textbf{Output}: Trusted client set $\mathcal C^{\rm{tr}}$\\
	\textbf{Initialization}: Defense period of MCD $T'$; Trusted client set $\mathcal C^{\rm{tr}}=\mathcal C$
	\begin{algorithmic}[1]
		\FOR{ $t=1,2,...,T$} 
		\IF{$t \% T'==0$ }
		\STATE $f=\lfloor \dfrac{t}{T'} \rfloor $
		\FOR{$\mathcal Q_{f,i}\in\mathcal Q_f$}
		\STATE Calculate model metrics $\boldsymbol{\bar{\theta}}_{f,i}^{(2)}$, $\bar{d}_{f,i}$ accoring to (15),(16)
		\STATE Use $(\boldsymbol{\bar{\theta}}_{f,i}^{(2)}, \bar{d}_{f,i})$ instead of $\mathcal Q_{f,i}$
		\ENDFOR
		\STATE Select baseline client set $\mathcal C^{\rm{base}}$ according to (18)(19)
		\STATE Run Algorithm 2: Client abnormality evaluation 
		\STATE Get outputted $\mathcal H_f$
		\STATE Calculate $h^{\rm{thr}}_f$ according to (23)
		\FOR{$c_{i}\in \mathcal C^{\rm{tr}}$}
		\IF{$h_{f,i}>h^{\rm{thr}}_f$}
		\STATE remove $c_i$ from $\mathcal C^{\rm{tr}}$
		\STATE remove $Q_{f,i}$ from $\mathcal Q_f$
		\ENDIF
		\ENDFOR
		\ENDIF
		\STATE Update $\mathcal C^{tr}, \mathcal Q_f$ 
		\ENDFOR 		
	\end{algorithmic} 
\end{algorithm}

\paragraph{\textbf{Step 2 (Baseline Value Calculation)}}MCD calculates baseline values $\boldsymbol{{\theta}}^{\rm{base}}_f$, ${d}^{\rm{base}}_f$ of above two metrics for anomaly detection. The calculation of baseline values is a crucial step in MCD as it directly influences the effectiveness of its judgment. 

Thus, MCD establishes a baseline client set $\mathcal C^{\rm{base}}$ for baseline values calculation, and the processes of selecting the baseline clients are as follows. Firstly, MCD maintains a first baseline client (for instance, the server acts as a such client to train a local model). This client is considered 100\% secure and immune to attacks, and its model metrics are denoted as $(\boldsymbol{\hat{\theta}}_{f}^{(2)}, \hat{d}_{f})$. The first baseline client can expand to multiple ones if possible. Secondly, MCD defines $C^{\rm{base}}$ as the set of baseline clients whose model metrics are within a neighborhood of those of the first baseline client. Two constants $\varOmega_1$, $\varOmega_2$ are used to define the size of the neighborhood:  
\begin{equation}
	\frac{\lVert \boldsymbol{\bar{\theta}}_{f,i}^{\left( 2 \right)}-\boldsymbol{\hat{\theta}}_{f}^{\left( 2 \right)} \rVert}{\hat{d}_{f}}<\varOmega_1 
\end{equation}
which represents the requirement for the centroid metrics, and 
\begin{equation}
	\frac{|\bar{d}_{f,i}-\hat{d}_{f}|}{\hat{d}_{f}}<\varOmega_2  
\end{equation}
which represents the requirement for the footprint metrics.

%The set composed of metrics that simultaneously satisfy equations (17) and (18) is denoted as $\mathcal U^{base}$. The set of clients associated with $\mathcal U^{base}$ is known as the baseline clients set $\mathcal C^{base}$. MCD calculates the baseline values using only the metrics from the set $\mathcal U^{base}$. 
After filtering out the baseline client set $\mathcal{C}^{\rm{base}}$ based on the above constraints, MCD utilizes the model metrics of the baseline clients to calculate the baseline values. The calculation of $\boldsymbol{{\theta}}^{\rm{base}}_f$ is as follows:
\begin{equation}
	\boldsymbol{{\theta}}^{\rm{base}}_f=\frac{\sum_{\boldsymbol{\bar{\theta}}_{f,i}^{(2)}\in \mathcal{C}^{\rm{base}}}{\boldsymbol{\bar{\theta}}_{f,i}^{(2)}}}{\left| \mathcal{C}^{\rm{base}} \right|}.
\end{equation}
Similarly, the calculation process of ${d}_{f}^{\rm{base}}$ is as follows:
\begin{equation}
	{d}_{f}^{\rm{base}}=\frac{\sum_{\bar{d}_{f,i}\in \mathcal{C}^{\rm{base}}}{\bar{d}_{f,i}}}{\left| \mathcal{C}^{\rm{base}} \right|}.
\end{equation}

So far, MCD has obtained the centroid baseline value $\boldsymbol{{\theta}}^{\rm{base}}_f$ and the footprint baseline value ${d}_{f}^{\rm{base}}$. 

\paragraph{\textbf{Step 3 (Client Abnormality Calculation)}}MCD uses the baseline values to calculate every trusted client abnormality separately. That is to calculate the degree of deviation of each client $c_i$'s metric pair $(\boldsymbol{\bar{\theta}}_{f,i}^{(2)}, \bar{d}_{f,i})$ from the baseline values. MCD quantifies it as $h_{f,i}$, and the calculation of $h_{f,i}$ is:
\begin{equation}
	h_{f,i}=\lambda _1\left(\frac{\lVert \boldsymbol{\bar{\theta}}_{f,i}^{(2)}-\boldsymbol{{\theta}}^{\rm{base}}_f \rVert}{{d}_{f}^{\rm{base}}}\right)+\lambda _2 \left(|{d}_{f}^{\rm{base}}-\bar{d}_{f,i}|\right).
\end{equation}
The client abnormality set of all trusted clients is denoted as $\mathcal H_f$. The first part of $h_{f,i}$ can be used to detect anomalies with the centroid metric, while the second part can be used to detect anomalies with the footprint metric. $\lambda _1$, $\lambda _2$ are weights and can be set as needed. The first half and the second half of Eq. (21) are not of the same order of magnitude. $\lambda _1$ and $\lambda _2$ are used to normalize the two parts. The steps for client abnormality evaluation are summarized in Algorithm 2.

\paragraph{\textbf{Step 4 (Malicious Client Identification)}}After obtaining all trusted client abnormality values, MCD finds a median value say $h^{\rm{med}}_f$. MCD then calculates an abnormality threshold $h^{\rm{thr}}_f$ for malicious client detection:
\begin{equation}
	h^{\rm{thr}}_f=\delta h^{\rm{med}}_f
\end{equation}
where parameter $\delta \in \left( 1,3 \right)$ controls the threshold. MCD then marks every client with a client abnormality value larger than the threshold as a malicious client. If MCD detects a malicious client, it will be removed from the trusted client set $\mathcal C^{\rm{tr} }$. The complete procedure of MCD, executed by the server, is outlined in Algorithm 3.
%If MCD detects a malicious client, MCD will remove the malicious client from $\mathcal C^{tr}$. The full steps of MCD performed by the server are summarized in Algorithm 3.

%and the FL system will no longer receive its local models in subsequent training rounds. 

%\begin{displaymath}
%	h_{t,i}>2\overline h_{t}
%\end{displaymath}
Overall, MCD is well-equipped to work against GAN-based data poisoning attacks as well as other mainstream ones. MCD run by the server takes advantage of a sufficient number of collected local models to conduct feature extraction so as to derive the two key metrics for model consistency evaluation. The centroid metric is useful for detecting most of the existing data poisoning attacks, such as label flipping, PoisonGAN, and noise superimposition, while the footprint metric is powerful for detecting GAN-based data poisoning attacks. As summarized in Tables II and III, MCD is more widely effective than existing defense methods. 

In addition, MCD can be implemented in a cost-effective manner. MCD operates without model collection overhead, since it directly reuses the local models already collected by the server for federated aggregation. Furthermore, on the premise of ensuring that the accuracy of judgment is not affected, some measures have been taken to ensure minimum impact on the server load: 1) MCD works with extracted metrics instead of original high-dimensional models, which avoids high space load on the server. 2) MCD uses a lightweight anomaly detection algorithm, which does not require complex computations. 3) MCD defends every period of tens of training rounds rather than every round, which avoids high computational load on the server. 

%\begin{figure}[h]
%	\centering
%	\includegraphics[width=\linewidth]{1000}
%	\caption{1000}
%\end{figure}
\section{Performance Evaluation}

In this section, we evaluate our presented VagueGAN and MCD through extensive experiments on multiple open real-world datasets.

\begin{figure}[h]
	\centering
	\includegraphics[width=\linewidth]{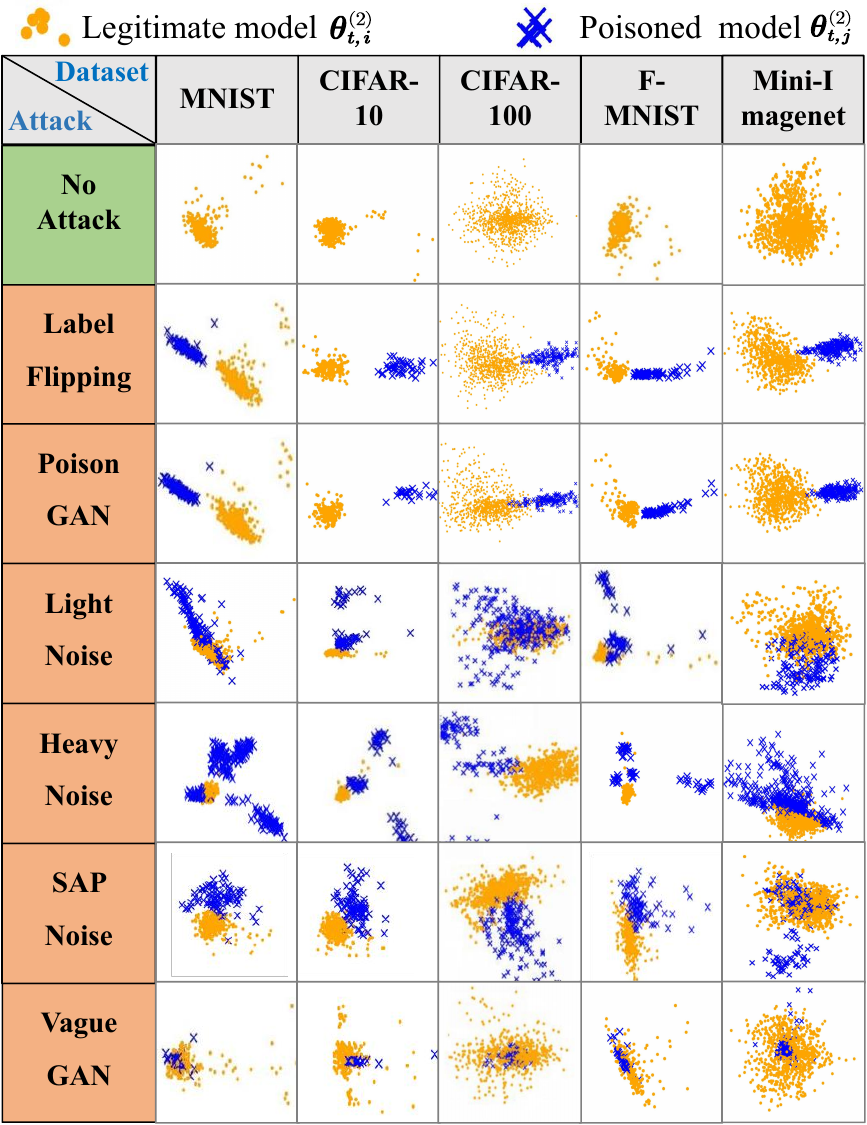}
	\caption{Two-dimensional local model visualization: VagueGAN shows higher stealthiness than the others due to a shorter distance between legitimate and poisoned models. }
\end{figure} 

\subsection{Experimental Settings}

\paragraph{\textbf{Datasets and Learning Task}} We take image data as an example to evaluate the poisoning capability of our VagueGAN, but the idea of ``vague data'' is not limited to this specific data type. We conduct our experiments on five typical real-world datasets: MNIST, Fashion-MNIST, CIFAR-10, CIFAR-100, and Mini-ImageNet. These datasets are commonly used in existing works. The learning task of our FL system is to train a deep learning model for image classification. In the experiments using MNIST and Fashion-MNIST, a CNN with two convolutional layers (16 and 32 channels, $5\times5$ kernels, padding 2), each followed by batch normalization, ReLU, and $2\times2$ max pooling, is used. A fully connected layer maps the $7\times7\times32$ features to 10 classes. In the experiments using CIFAR-10, a CNN with six convolutional layers (using $3\times3$ kernels and padding 1), grouped into three blocks with 32, 64, and 128 channels respectively, each block followed by batch normalization, ReLU activation, and $2\times2$ max pooling, is used. Two fully connected layers map the $4\times4\times128$ features to 10 output classes. The model used for CIFAR-100 shares the same depth but has twice the width (i.e., number of channels) compared to that used for CIFAR-10, in order to better handle the higher number of classes. In the experiments using Mini-ImageNet, we employ a CNN with three convolutional blocks having 64, 128, and 256 channels, respectively. Uniquely, this architecture incorporates an adaptive average pooling layer to standardize feature dimensions to $4\times4$, followed by a dropout layer and two fully connected layers for 100-class classification.

\begin{table*}[htbp]
	\caption{Test accuracy under various data poisoning attacks (\%) on different datasets: The 0\% column is for the case without any attack, while the other columns are for the cases with different proportions of malicious clients in the FL system.}
	\begin{center}
		\setlength{\tabcolsep}{3.710mm}
		\begin{tabular*}{\linewidth}{ccccccccccc}
			\hline
			\multicolumn{1}{c|}{Client data distribution} &
			\multicolumn{5}{c|}{IID} &
			\multicolumn{5}{c}{non-IID} \\ \hline
			\multicolumn{1}{c|}{} &
			\multicolumn{10}{c}{Malicious Clients Percentage $\alpha$} \\ \cline{2-11} 
			\multicolumn{1}{c|}{\multirow{-2}{*}{Poisoning Attack}} &
			\multicolumn{1}{c|}{0\%} &
			\multicolumn{1}{c|}{5\%} &
			\multicolumn{1}{c|}{10\%} &
			\multicolumn{1}{c|}{20\%} &
			\multicolumn{1}{c|}{30\%} &
			\multicolumn{1}{c|}{0\%} &
			\multicolumn{1}{c|}{5\%} &
			\multicolumn{1}{c|}{10\%} &
			\multicolumn{1}{c|}{20\%} &
			30\% \\ \hline
			\multicolumn{11}{c}{MNIST} \\ \hline
			\multicolumn{1}{c|}{Label Flipping} &
			\multicolumn{1}{c|}{98.73} &
			\multicolumn{1}{c|}{98.65} &
			\multicolumn{1}{c|}{98.48} &
			\multicolumn{1}{c|}{98.32} &
			\multicolumn{1}{c|}{98.11} &
			\multicolumn{1}{c|}{98.39} &
			\multicolumn{1}{c|}{98.32} &
			\multicolumn{1}{c|}{98.08} &
			\multicolumn{1}{c|}{97.92} &
			97.59 \\ \cline{1-1}
			\multicolumn{1}{c|}{Label Flipping+PoisonGAN} &
			\multicolumn{1}{c|}{98.73} &
			\multicolumn{1}{c|}{98.64} &
			\multicolumn{1}{c|}{98.43} &
			\multicolumn{1}{c|}{98.21} &
			\multicolumn{1}{c|}{98.04} &
			\multicolumn{1}{c|}{98.39} &
			\multicolumn{1}{c|}{98.3} &
			\multicolumn{1}{c|}{\cellcolor[HTML]{FFFC9E}97.91} &
			\multicolumn{1}{c|}{97.88} &
			97.44 \\ \cline{1-1}
			\multicolumn{1}{c|}{Light Noise Superimposition} &
			\multicolumn{1}{c|}{98.73} &
			\multicolumn{1}{c|}{98.66} &
			\multicolumn{1}{c|}{98.57} &
			\multicolumn{1}{c|}{98.41} &
			\multicolumn{1}{c|}{98.31} &
			\multicolumn{1}{c|}{98.39} &
			\multicolumn{1}{c|}{98.37} &
			\multicolumn{1}{c|}{98.15} &
			\multicolumn{1}{c|}{98.02} &
			97.74 \\ \cline{1-1}
			\multicolumn{1}{c|}{Heavy Noise Superimposition} &
			\multicolumn{1}{c|}{98.73} &
			\multicolumn{1}{c|}{98.63} &
			\multicolumn{1}{c|}{98.49} &
			\multicolumn{1}{c|}{98.37} &
			\multicolumn{1}{c|}{98.11} &
			\multicolumn{1}{c|}{98.39} &
			\multicolumn{1}{c|}{98.36} &
			\multicolumn{1}{c|}{98.03} &
			\multicolumn{1}{c|}{97.89} &
			97.61 \\ \cline{1-1}
			\multicolumn{1}{c|}{SAP Noise Superimposition} &
			\multicolumn{1}{c|}{98.73} &
			\multicolumn{1}{c|}{98.70} &
			\multicolumn{1}{c|}{98.62} &
			\multicolumn{1}{c|}{98.56} &
			\multicolumn{1}{c|}{98.43} &
			\multicolumn{1}{c|}{98.39} &
			\multicolumn{1}{c|}{98.37} &
			\multicolumn{1}{c|}{98.19} &
			\multicolumn{1}{c|}{98.04} &
			97.82 \\ \cline{1-1}
			\multicolumn{1}{c|}{Cosine Noise Superimposition} &
			\multicolumn{1}{c|}{98.73} &
			\multicolumn{1}{c|}{98.69} &
			\multicolumn{1}{c|}{98.60} &
			\multicolumn{1}{c|}{98.45} &
			\multicolumn{1}{c|}{98.24} &
			\multicolumn{1}{c|}{98.39} &
			\multicolumn{1}{c|}{98.39} &
			\multicolumn{1}{c|}{98.15} &
			\multicolumn{1}{c|}{98.00} &
			97.18 \\ \cline{1-1}
			\multicolumn{1}{c|}{VagueGAN} &
			\multicolumn{1}{c|}{98.73} &
			\multicolumn{1}{c|}{\cellcolor[HTML]{FFFC9E}98.61} &
			\multicolumn{1}{c|}{\cellcolor[HTML]{FFFC9E}98.35} &
			\multicolumn{1}{c|}{\cellcolor[HTML]{FFFC9E}97.84} &
			\multicolumn{1}{c|}{\cellcolor[HTML]{FFFC9E}97.46} &
			\multicolumn{1}{c|}{98.39} &
			\multicolumn{1}{c|}{\cellcolor[HTML]{FFFC9E}98.13} &
			\multicolumn{1}{c|}{97.94} &
			\multicolumn{1}{c|}{\cellcolor[HTML]{FFFC9E}97.77} &
			\cellcolor[HTML]{FFFC9E}97.37 \\ \hline
			\multicolumn{11}{c}{Fashion-MNIST} \\ \hline
			\multicolumn{1}{c|}{Label Flipping} &
			\multicolumn{1}{c|}{87.97} &
			\multicolumn{1}{c|}{87.73} &
			\multicolumn{1}{c|}{87.20} &
			\multicolumn{1}{c|}{86.01} &
			\multicolumn{1}{c|}{85.48} &
			\multicolumn{1}{c|}{87.25} &
			\multicolumn{1}{c|}{86.73} &
			\multicolumn{1}{c|}{86.18} &
			\multicolumn{1}{c|}{85.31} &
			84.36 \\ \cline{1-1}
			\multicolumn{1}{c|}{Label Flipping+PoisonGAN} &
			\multicolumn{1}{c|}{87.97} &
			\multicolumn{1}{c|}{87.59} &
			\multicolumn{1}{c|}{87.17} &
			\multicolumn{1}{c|}{\cellcolor[HTML]{FFFC9E}85.77} &
			\multicolumn{1}{c|}{85.09} &
			\multicolumn{1}{c|}{87.25} &
			\multicolumn{1}{c|}{\cellcolor[HTML]{FFFC9E}86.59} &
			\multicolumn{1}{c|}{86.01} &
			\multicolumn{1}{c|}{85.04} &
			84.08 \\ \cline{1-1}
			\multicolumn{1}{c|}{Light Noise Superimposition} &
			\multicolumn{1}{c|}{87.97} &
			\multicolumn{1}{c|}{87.81} &
			\multicolumn{1}{c|}{87.58} &
			\multicolumn{1}{c|}{86.92} &
			\multicolumn{1}{c|}{86.57} &
			\multicolumn{1}{c|}{87.25} &
			\multicolumn{1}{c|}{87.08} &
			\multicolumn{1}{c|}{86.82} &
			\multicolumn{1}{c|}{86.29} &
			85.72 \\ \cline{1-1}
			\multicolumn{1}{c|}{Heavy Noise Superimposition} &
			\multicolumn{1}{c|}{87.97} &
			\multicolumn{1}{c|}{87.71} &
			\multicolumn{1}{c|}{87.15} &
			\multicolumn{1}{c|}{86.16} &
			\multicolumn{1}{c|}{85.56} &
			\multicolumn{1}{c|}{87.25} &
			\multicolumn{1}{c|}{86.94} &
			\multicolumn{1}{c|}{86.57} &
			\multicolumn{1}{c|}{85.83} &
			85.33 \\ \cline{1-1}
			\multicolumn{1}{c|}{SAP Noise Superimposition} &
			\multicolumn{1}{c|}{87.97} &
			\multicolumn{1}{c|}{87.80} &
			\multicolumn{1}{c|}{87.52} &
			\multicolumn{1}{c|}{86.99} &
			\multicolumn{1}{c|}{86.43} &
			\multicolumn{1}{c|}{87.25} &
			\multicolumn{1}{c|}{87.12} &
			\multicolumn{1}{c|}{86.77} &
			\multicolumn{1}{c|}{86.32} &
			85.94 \\ \cline{1-1}
			\multicolumn{1}{c|}{Cosine Noise Superimposition} &
			\multicolumn{1}{c|}{87.97} &
			\multicolumn{1}{c|}{87.78} &
			\multicolumn{1}{c|}{87.48} &
			\multicolumn{1}{c|}{86.79} &
			\multicolumn{1}{c|}{86.50} &
			\multicolumn{1}{c|}{87.25} &
			\multicolumn{1}{c|}{87.08} &
			\multicolumn{1}{c|}{86.65} &
			\multicolumn{1}{c|}{86.40} &
			85.71 \\ \cline{1-1}
			\multicolumn{1}{c|}{VagueGAN} &
			\multicolumn{1}{c|}{87.97} &
			\multicolumn{1}{c|}{\cellcolor[HTML]{FFFC9E}87.34} &
			\multicolumn{1}{c|}{\cellcolor[HTML]{FFFC9E}86.79} &
			\multicolumn{1}{c|}{85.87} &
			\multicolumn{1}{c|}{\cellcolor[HTML]{FFFC9E}84.77} &
			\multicolumn{1}{c|}{87.25} &
			\multicolumn{1}{c|}{86.64} &
			\multicolumn{1}{c|}{\cellcolor[HTML]{FFFC9E}85.84} &
			\multicolumn{1}{c|}{\cellcolor[HTML]{FFFC9E}84.91} &
			\cellcolor[HTML]{FFFC9E}84.02 \\ \hline
			\multicolumn{11}{c}{CIFAR-10} \\ \hline
			\multicolumn{1}{c|}{Label Flipping} &
			\multicolumn{1}{c|}{76.28} &
			\multicolumn{1}{c|}{75.69} &
			\multicolumn{1}{c|}{74.99} &
			\multicolumn{1}{c|}{73.86} &
			\multicolumn{1}{c|}{72.85} &
			\multicolumn{1}{c|}{73.61} &
			\multicolumn{1}{c|}{72.32} &
			\multicolumn{1}{c|}{71.23} &
			\multicolumn{1}{c|}{70.18} &
			68.78 \\ \cline{1-1}
			\multicolumn{1}{c|}{Label Flipping+PoisonGAN} &
			\multicolumn{1}{c|}{76.28} &
			\multicolumn{1}{c|}{75.62} &
			\multicolumn{1}{c|}{74.84} &
			\multicolumn{1}{c|}{73.77} &
			\multicolumn{1}{c|}{72.66} &
			\multicolumn{1}{c|}{73.61} &
			\multicolumn{1}{c|}{72.16} &
			\multicolumn{1}{c|}{71.06} &
			\multicolumn{1}{c|}{70.04} &
			68.66 \\ \cline{1-1}
			\multicolumn{1}{c|}{Light Noise Superimposition} &
			\multicolumn{1}{c|}{76.28} &
			\multicolumn{1}{c|}{76.03} &
			\multicolumn{1}{c|}{75.71} &
			\multicolumn{1}{c|}{74.93} &
			\multicolumn{1}{c|}{74.07} &
			\multicolumn{1}{c|}{73.61} &
			\multicolumn{1}{c|}{72.94} &
			\multicolumn{1}{c|}{72.03} &
			\multicolumn{1}{c|}{71.26} &
			70.53 \\ \cline{1-1}
			\multicolumn{1}{c|}{Heavy Noise Superimposition} &
			\multicolumn{1}{c|}{76.28} &
			\multicolumn{1}{c|}{75.88} &
			\multicolumn{1}{c|}{75.13} &
			\multicolumn{1}{c|}{74.24} &
			\multicolumn{1}{c|}{73.11} &
			\multicolumn{1}{c|}{73.61} &
			\multicolumn{1}{c|}{72.82} &
			\multicolumn{1}{c|}{71.88} &
			\multicolumn{1}{c|}{70.73} &
			69.94 \\ \cline{1-1}
			\multicolumn{1}{c|}{SAP Noise Superimposition} &
			\multicolumn{1}{c|}{76.28} &
			\multicolumn{1}{c|}{76.12} &
			\multicolumn{1}{c|}{75.73} &
			\multicolumn{1}{c|}{74.71} &
			\multicolumn{1}{c|}{73.96} &
			\multicolumn{1}{c|}{73.61} &
			\multicolumn{1}{c|}{72.99} &
			\multicolumn{1}{c|}{71.97} &
			\multicolumn{1}{c|}{71.22} &
			70.38 \\ \cline{1-1}
			\multicolumn{1}{c|}{Cosine Noise Superimposition} &
			\multicolumn{1}{c|}{76.28} &
			\multicolumn{1}{c|}{76.10} &
			\multicolumn{1}{c|}{75.67} &
			\multicolumn{1}{c|}{74.54} &
			\multicolumn{1}{c|}{73.87} &
			\multicolumn{1}{c|}{73.61} &
			\multicolumn{1}{c|}{72.88} &
			\multicolumn{1}{c|}{71.89} &
			\multicolumn{1}{c|}{71.36} &
			70.12 \\ \cline{1-1}
			\multicolumn{1}{c|}{VagueGAN} &
			\multicolumn{1}{c|}{76.28} &
			\multicolumn{1}{c|}{\cellcolor[HTML]{FFFC9E}74.94} &
			\multicolumn{1}{c|}{\cellcolor[HTML]{FFFC9E}74.02} &
			\multicolumn{1}{c|}{\cellcolor[HTML]{FFFC9E}73.16} &
			\multicolumn{1}{c|}{\cellcolor[HTML]{FFFC9E}71.93} &
			\multicolumn{1}{c|}{73.61} &
			\multicolumn{1}{c|}{\cellcolor[HTML]{FFFC9E}71.96} &
			\multicolumn{1}{c|}{\cellcolor[HTML]{FFFC9E}70.64} &
			\multicolumn{1}{c|}{\cellcolor[HTML]{FFFC9E}69.55} &
			\cellcolor[HTML]{FFFC9E}67.79 \\ \hline
			
			\multicolumn{11}{c}{CIFAR-100} \\ \hline
			\multicolumn{1}{c|}{Label Flipping(x10)} &
			\multicolumn{1}{c|}{55.71} &
			\multicolumn{1}{c|}{53.03} &
			\multicolumn{1}{c|}{52.52} &
			\multicolumn{1}{c|}{49.25} &
			\multicolumn{1}{c|}{46.68} &
			\multicolumn{1}{c|}{53.43} &
			\multicolumn{1}{c|}{52.07} &
			\multicolumn{1}{c|}{49.26} &
			\multicolumn{1}{c|}{46.78} &
			41.39 \\ \cline{1-1}
			\multicolumn{1}{c|}{Label Flipping+PoisonGAN} &
			\multicolumn{1}{c|}{55.71} &
			\multicolumn{1}{c|}{52.83} &
			\multicolumn{1}{c|}{52.10} &
			\multicolumn{1}{c|}{48.39} &
			\multicolumn{1}{c|}{46.93} &
			\multicolumn{1}{c|}{53.43} &
			\multicolumn{1}{c|}{51.66} &
			\multicolumn{1}{c|}{49.37} &
			\multicolumn{1}{c|}{46.68} &
			40.03 \\ \cline{1-1}
			\multicolumn{1}{c|}{Light Noise Superimposition} &
			\multicolumn{1}{c|}{55.71} &
			\multicolumn{1}{c|}{54.03} &
			\multicolumn{1}{c|}{53.67} &
			\multicolumn{1}{c|}{51.08} &
			\multicolumn{1}{c|}{49.95} &
			\multicolumn{1}{c|}{53.43} &
			\multicolumn{1}{c|}{52.92} &
			\multicolumn{1}{c|}{50.97} &
			\multicolumn{1}{c|}{48.50} &
			45.95 \\ \cline{1-1}
			\multicolumn{1}{c|}{Heavy Noise Superimposition} &
			\multicolumn{1}{c|}{55.71} &
			\multicolumn{1}{c|}{53.49} &
			\multicolumn{1}{c|}{52.45} &
			\multicolumn{1}{c|}{50.23} &
			\multicolumn{1}{c|}{48.58} &
			\multicolumn{1}{c|}{53.43} &
			\multicolumn{1}{c|}{51.29} &
			\multicolumn{1}{c|}{\cellcolor[HTML]{FFFC9E}48.62} &
			\multicolumn{1}{c|}{45.05} &
			42.98 \\ \cline{1-1}
			\multicolumn{1}{c|}{SAP Noise Superimposition} &
			\multicolumn{1}{c|}{55.71} &
			\multicolumn{1}{c|}{54.12} &
			\multicolumn{1}{c|}{53.80} &
			\multicolumn{1}{c|}{51.21} &
			\multicolumn{1}{c|}{49.63} &
			\multicolumn{1}{c|}{53.43} &
			\multicolumn{1}{c|}{51.72} &
			\multicolumn{1}{c|}{49.97} &
			\multicolumn{1}{c|}{48.43} &
			44.68 \\ \cline{1-1}
			\multicolumn{1}{c|}{Cosine Noise Superimposition} &
			\multicolumn{1}{c|}{55.71} &
			\multicolumn{1}{c|}{54.37} &
			\multicolumn{1}{c|}{53.53} &
			\multicolumn{1}{c|}{51.17} &
			\multicolumn{1}{c|}{49.15} &
			\multicolumn{1}{c|}{53.43} &
			\multicolumn{1}{c|}{51.67} &
			\multicolumn{1}{c|}{49.61} &
			\multicolumn{1}{c|}{48.48} &
			43.56 \\ \cline{1-1}
			\multicolumn{1}{c|}{VagueGAN} &
			\multicolumn{1}{c|}{55.71} &
			\multicolumn{1}{c|}{\cellcolor[HTML]{FFFC9E}51.53} &
			\multicolumn{1}{c|}{\cellcolor[HTML]{FFFC9E}48.30} &
			\multicolumn{1}{c|}{\cellcolor[HTML]{FFFC9E}42.35} &
			\multicolumn{1}{c|}{\cellcolor[HTML]{FFFC9E}39.38} &
			\multicolumn{1}{c|}{53.43} &
			\multicolumn{1}{c|}{\cellcolor[HTML]{FFFC9E}51.04} &
			\multicolumn{1}{c|}{48.96} &
			\multicolumn{1}{c|}{\cellcolor[HTML]{FFFC9E}41.02} &
			\cellcolor[HTML]{FFFC9E}35.11 \\ \hline
			
			\multicolumn{11}{c}{Mini-ImageNet} \\ \hline
			\multicolumn{1}{c|}{Label Flipping(x10)} &
			\multicolumn{1}{c|}{49.25} &
			\multicolumn{1}{c|}{48.83} &
			\multicolumn{1}{c|}{47.19} &
			\multicolumn{1}{c|}{44.63} &
			\multicolumn{1}{c|}{42.01} &
			\multicolumn{1}{c|}{42.94} &
			\multicolumn{1}{c|}{41.30} &
			\multicolumn{1}{c|}{39.22} &
			\multicolumn{1}{c|}{35.76} &
			31.83 \\ \cline{1-1}
			\multicolumn{1}{c|}{Label Flipping+PoisonGAN} &
			\multicolumn{1}{c|}{49.25} &
			\multicolumn{1}{c|}{48.30} &
			\multicolumn{1}{c|}{46.92} &
			\multicolumn{1}{c|}{43.75} &
			\multicolumn{1}{c|}{41.40} &
			\multicolumn{1}{c|}{42.94} &
			\multicolumn{1}{c|}{40.89} &
			\multicolumn{1}{c|}{37.08} &
			\multicolumn{1}{c|}{33.81} &
			30.42 \\ \cline{1-1}
			\multicolumn{1}{c|}{Light Noise Superimposition} &
			\multicolumn{1}{c|}{49.25} &
			\multicolumn{1}{c|}{47.91} &
			\multicolumn{1}{c|}{46.12} &
			\multicolumn{1}{c|}{43.28} &
			\multicolumn{1}{c|}{40.02} &
			\multicolumn{1}{c|}{42.94} &
			\multicolumn{1}{c|}{40.88} &
			\multicolumn{1}{c|}{36.09} &
			\multicolumn{1}{c|}{34.53} &
			30.61 \\ \cline{1-1}
			\multicolumn{1}{c|}{Heavy Noise Superimposition} &
			\multicolumn{1}{c|}{49.25} &
			\multicolumn{1}{c|}{46.98} &
			\multicolumn{1}{c|}{44.16} &
			\multicolumn{1}{c|}{39.81} &
			\multicolumn{1}{c|}{36.17} &
			\multicolumn{1}{c|}{42.94} &
			\multicolumn{1}{c|}{38.29} &
			\multicolumn{1}{c|}{34.41} &
			\multicolumn{1}{c|}{30.71} &
			25.67 \\ \cline{1-1}
			\multicolumn{1}{c|}{SAP Noise Superimposition} &
			\multicolumn{1}{c|}{49.25} &
			\multicolumn{1}{c|}{46.59} &
			\multicolumn{1}{c|}{45.59} &
			\multicolumn{1}{c|}{41.39} &
			\multicolumn{1}{c|}{38.22} &
			\multicolumn{1}{c|}{42.94} &
			\multicolumn{1}{c|}{40.00} &
			\multicolumn{1}{c|}{36.67} &
			\multicolumn{1}{c|}{31.99} &
			27.30 \\ \cline{1-1}
			\multicolumn{1}{c|}{Cosine Noise Superimposition} &
			\multicolumn{1}{c|}{49.25} &
			\multicolumn{1}{c|}{47.65} &
			\multicolumn{1}{c|}{45.63} &
			\multicolumn{1}{c|}{41.53} &
			\multicolumn{1}{c|}{39.71} &
			\multicolumn{1}{c|}{42.94} &
			\multicolumn{1}{c|}{40.26} &
			\multicolumn{1}{c|}{35.74} &
			\multicolumn{1}{c|}{31.24} &
			26.81 \\ \cline{1-1}
			\multicolumn{1}{c|}{VagueGAN} &
			\multicolumn{1}{c|}{49.25} &
			\multicolumn{1}{c|}{\cellcolor[HTML]{FFFC9E}46.12} &
			\multicolumn{1}{c|}{\cellcolor[HTML]{FFFC9E}41.16} &
			\multicolumn{1}{c|}{\cellcolor[HTML]{FFFC9E}35.22} &
			\multicolumn{1}{c|}{\cellcolor[HTML]{FFFC9E}29.35} &
			\multicolumn{1}{c|}{42.94} &
			\multicolumn{1}{c|}{\cellcolor[HTML]{FFFC9E}37.33} &
			\multicolumn{1}{c|}{\cellcolor[HTML]{FFFC9E}32.24} &
			\multicolumn{1}{c|}{\cellcolor[HTML]{FFFC9E}27.06} &
			\cellcolor[HTML]{FFFC9E}19.76 \\ \hline
		\end{tabular*}
		\label{tab2}
	\end{center}
	\vspace{-15pt}
\end{table*}

\begin{table*}[ht]
	\centering
	
	\caption{Detection results of MCD under three attacks: Red indicates malicious clients; green indicates benign clients.}
	\label{1-3}
	\begin{tabular}{c|>{\centering\arraybackslash}p{2cm}>{\centering\arraybackslash}p{1.2cm} >{\centering\arraybackslash}p{1.2cm}|>{\centering\arraybackslash}p{2cm} >{\centering\arraybackslash}p{1.2cm} >{\centering\arraybackslash}p{1.2cm}|>{\centering\arraybackslash}p{2cm} >{\centering\arraybackslash}p{1.2cm} >{\centering\arraybackslash}p{1.2cm}}
		
		\hline
		\multirow{2}{*}{Scenario} & \multicolumn{3}{c|}{Label flipping}                   & \multicolumn{3}{c|}{PoisonGAN}                       & \multicolumn{3}{c}{Light noise superimposition}     \\[0.1cm]
		& \multicolumn{3}{c|}{$\boldsymbol{ {\theta}}^{base}_f$=(-5.31,-0.62), ${d}_{f}^{base}$=3.32} & \multicolumn{3}{c|}{$\boldsymbol{{\theta}}^{base}_f$=(-4.57,0.44), ${d}_{f}^{base}$= 4.15} & \multicolumn{3}{c}{$\boldsymbol{{\theta}}^{base}_f$=(-1.15, 0.25), ${d}_{f}^{base}$=2.72} \\ \hline
		i                         & $\boldsymbol{\bar{\theta}}_{f,i}^{(2)}$                       & $\bar{d}_{f,i}$            & $h_{f,i}$            & $\boldsymbol{\bar{\theta}}_{f,i}^{(2)}$                      & $\bar{d}_{f,i}$           & $h_{f,i}$            & $\boldsymbol{\bar{\theta}}_{f,i}^{(2)}$                       & $\bar{d}_{f,i}$          & $h_{f,i}$            \\ \hline
		0                         & (-6.22,4.26)           & 4.09          & 1.49         & \cellcolor[HTML]{FFCCC9} (19.61,-0.24) & 3.68 & \cellcolor[HTML]{FFCCC9} 6.77         & \cellcolor[HTML]{FFCCC9} (3.35,10.36)           & 3.89        & \cellcolor[HTML]{FFCCC9} 6.41         \\
		1                         & \cellcolor[HTML]{FFCCC9} (21.87,-0.58) &  3.07 & \cellcolor[HTML]{FFCCC9} 8.74 & (-4.49,-0.33) & 5.33 & 2.55 &\cellcolor[HTML]{9AFF99} (-0.40,1.20) & \cellcolor[HTML]{9AFF99}2.53 & 0.82 \\
		2                         & (-1.86,-9.85)          & 3.79          & 2.97         & \cellcolor[HTML]{FFCCC9} (21.47,-0.52)          &  3.06         & \cellcolor[HTML]{FFCCC9} 8.46         & (2.66,2.45)            & 2.91        & 2            \\
		3                         & (-2.67,11.74)          & 3.13          & 4.18         & (-4.73,-2.7)           & 4.46         & 1.38         & \cellcolor[HTML]{FFCCC9} (8.81,2.52) &  5.08 & \cellcolor[HTML]{FFCCC9} 8.48 \\
		4                         & (-3.58,3.63)           & 4.44          & 1.38         & (-2.09,3.05)           & 4.62         & 1.81         & (-1.10,0.05)           & 1.71        & 2.1          \\
		5                         & \cellcolor[HTML]{9AFF99}(-4.64,-3.19)          &\cellcolor[HTML]{9AFF99} 3.09          & 1.26         & (-5.24,-1.12)          & 5.1          & 2.31         & (-3.96,-2.02)          & 2.02        & 2.73         \\
		6                         & (-7.15,0.66)           & 3.43          & 0.67         & (-3.13,1.56)           & 4.02         & 0.7          & \cellcolor[HTML]{FFCCC9} (13.32,-26.22) &  4.04 & \cellcolor[HTML]{FFCCC9} 13.71 \\
		7                         & (-6.87,-3.16)          & 2.85          & 1.84         & (-3.22,6.41)           & 4.09         & 1.59         & (-3.05,0.80)           & 3.88        & 3.05         \\
		8                         & (-4.59,-2.59)          & 3.2           & 0.87         & (-6.38,4.06)           & 4.48         & 1.64         & (1.40,-3.55)           & 2.37        & 2.38         \\
		9                         & \cellcolor[HTML]{FFCCC9} (23.13,5.61) & 3.14 & \cellcolor[HTML]{FFCCC9} 9.13 & (-5.93,0.64) & 3.6 & 1.43 & (-1.07,1.17) & 3.9 & 2.7 \\
		10                        & \cellcolor[HTML]{FFCCC9} (21.51,-4.57) &  4.45 &\cellcolor[HTML]{FFCCC9}  8.17         & \cellcolor[HTML]{FFCCC9} (22.25,-0.53) &  3.31 & \cellcolor[HTML]{FFCCC9} 8.15 & (-4.34,3.92) & 2.19 & 2.81 \\
		11                        & (-5.27,1.53)           & 3.92          & 0.65         & \cellcolor[HTML]{FFCCC9} (21.66,-0.57) & 2.98 & \cellcolor[HTML]{FFCCC9} 8.67 & (-3.22,1.31) & 1.68 & 2.94 \\
		12                        & (-3.44,-7.33)          & 3.03          & 2.67         & (-3.95,2.86)           & 4.03         & 0.84         & (-0.76,-1.64)          & 2.17        & 1.81         \\
		13                        & (-5.83,2.51)           & 3.3           & 0.99         & (-5.07,1.43)           & 4.25         & 0.47         & (-1.50,0.18)           & 1.37        & 2.83         \\
		14                        & (-4.82,-1.70)          & 2.84          & 1.34         & (-6.55,4.28)           & 4.69         & 1.94         & (-2.14,2.11)           & 2.7         & 0.82         \\
		15                        & (-7.19,-6.73)          & 2.78          & 3.26         & \cellcolor[HTML]{9AFF99}(-5.73,-1.12)          & \cellcolor[HTML]{9AFF99}4.32         & 0.81         & (0.37,-1.00)           & 2.42        & 1.32         \\
		16                        & (-8.05,-6.00)          & 2.82          & 2.82         & (-3.94,-2.67)          & 4.01         & 1.04         & (2.60,1.76)            & 3.02        & 2.09         \\
		17                        & (-4.93,0.06)           & 2.97          & 0.93         & (-3.32,-5.79)          & 4.22         & 1.71         & (-1.63,-3.89)          & 2.3         & 2.35         \\
		18                        & (-7.89,0.25)           & 2.25          & 2.95         & (-4.29,-5.14)          & 4.12         & 1.56         & (-2.33,1.20)           & 1.12        & 1.12         \\
		19                        & \cellcolor[HTML]{FFCCC9} (22.01,-0.43) & 3.43 & \cellcolor[HTML]{FFCCC9} 8.23 & (-4.98,1.64) & 4.65 & 1.31 & \cellcolor[HTML]{FFCCC9} (4.27,12.33) &  3.75 & \cellcolor[HTML]{FFCCC9} 6.89 \\
		\hline
	\end{tabular}
	\vspace{-5pt}
\end{table*}

\begin{table*}[ht]
	\centering
	\caption{Detection results of MCD under another three attacks.}
	\label{4-6}
	\begin{tabular}{c|>{\centering\arraybackslash}p{2cm}>{\centering\arraybackslash}p{1.2cm} >{\centering\arraybackslash}p{1.2cm}|>{\centering\arraybackslash}p{2cm} >{\centering\arraybackslash}p{1.2cm} >{\centering\arraybackslash}p{1.2cm}|>{\centering\arraybackslash}p{2cm} >{\centering\arraybackslash}p{1.2cm} >{\centering\arraybackslash}p{1.2cm}}
		\hline
		\multirow{2}{*}{Scenario}  & \multicolumn{3}{c|}{Heavy noise superimposition}    & \multicolumn{3}{c|}{SAP noise superimposition}      & \multicolumn{3}{c}{VagueGAN}                      \\[0.1cm]
		& \multicolumn{3}{c|}{$\boldsymbol{{\theta}}^{base}_f$=(-0.51,-0.62),${d}_{f}^{base}$=0.82} & \multicolumn{3}{c|}{$\boldsymbol{{\theta}}^{base}_f$=(-0.75,-3.70),${d}_{f}^{base}$=1.50} & \multicolumn{3}{c}{$\boldsymbol{{\theta}}^{base}_f$=(-0.05,0.02),${d}_{f}^{base}$=5.55} \\ \hline
		i                         & $\boldsymbol{\bar{\theta}}_{f,i}^{(2)}$                     & $\bar{d}_{f,i}$          & $h_{f,i}$            & $\boldsymbol{\bar{\theta}}_{f,i}^{(2)}$                      & $\bar{d}_{f,i}$           & $h_{f,i}$           & $\boldsymbol{\bar{\theta}}_{f,i}^{(2)}$                     & $\bar{d}_{f,i}$          & $h_{f,i}$           \\ \hline
		0                         & (-0.81,-1.72)          & 0.73        & 1.48         & (-2.32,-2.46)          & 1.9          & 2.8         & (-2.52,0.93)          & \cellcolor[HTML]{FFCCC9}0.92        &\cellcolor[HTML]{FFCCC9} 8.92        \\
		1                         & (0.01,-0.86)           & 0.75        & 0.77         & \cellcolor[HTML]{FFCCC9}(-13.38,14.01)         & \cellcolor[HTML]{FFCCC9}3.06         & \cellcolor[HTML]{FFCCC9}24.82       & (-1.95,1.07)          & \cellcolor[HTML]{FFCCC9}1.01        & \cellcolor[HTML]{FFCCC9}8.64        \\
		2                         & (0.40,0.01)            & 0.76        & 1.41         & (0.28,-6.14)           & 1.38         & 2.88        & (-1.62,0.42)          & 6.13        & 0.32        \\
		3                         & (-0.30,0.09)           & 0.93        & 1.01         & (1.23,-5.05)           & 1.41         & 2.57        & (-2.77,4.45)          & 4.22        & 2.81        \\
		4                         & \cellcolor[HTML]{FFCCC9}(32.10,-10.60)         & \cellcolor[HTML]{FFCCC9}6.22        & \cellcolor[HTML]{FFCCC9}46.98        & (-1.61,-0.55)          & 2.42         & 5.1         & (-0.19,-2.55)         & 4.31        & 2.12        \\
		5                         & (0.83,-1.00)           & 0.69        & 1.83         & (-3.28,-2.26)          & 1.62         & 3.15        & (0.01,-0.88)          & 3.29        & 3.49        \\
		6                         & (-0.87,-0.68)          & 0.8         & 0.46         & (-1.55,-2.85)          & 1.56         & 1.28        & (-0.40,-1.14)         & 3.68        & 3.11        \\
		7                         & \cellcolor[HTML]{FFCCC9}(-22.62,-19.32)        & \cellcolor[HTML]{FFCCC9}6.18        & \cellcolor[HTML]{FFCCC9}40.68        & (-0.41,-3.78)          & 1.2          & 0.9         & (-2.45,-3.76)         & 4.3         & 2.51        \\
		8                         & \cellcolor[HTML]{FFCCC9}(-7.12,28.24)          & \cellcolor[HTML]{FFCCC9}5.86        & \cellcolor[HTML]{FFCCC9}41.15        & \cellcolor[HTML]{FFCCC9}(-12.11,8.79)          &\cellcolor[HTML]{FFCCC9} 2.6          &\cellcolor[HTML]{FFCCC9} 19.09       & (-2.01,0.32)          & 4.54        & 1.35        \\
		9                         & (-1.79,-0.92)          & 0.78        & 1.64         & (-1.81,-3.06)          & 1.12         & 1.99        & (3.34,-0.24)          & 6.45        & 0.66        \\
		10                        & (0.11,-0.87)           & 1.03        & 1.03         & \cellcolor[HTML]{FFCCC9}(35.53,9.16)           & \cellcolor[HTML]{FFCCC9}4.52         & \cellcolor[HTML]{FFCCC9}44.44       & (-1.66,1.02)          & \cellcolor[HTML]{FFCCC9}1.35        & \cellcolor[HTML]{FFCCC9}7.9         \\
		11                        & (-0.88,-0.56)          & 0.85        & 0.48         & (2.88,-6.84)           & 1.01         & 5.77        & (-1.03,1.21)          & 6.59        & 0.3         \\
		12                        & \cellcolor[HTML]{FFCCC9}(4.70,14.92)           & \cellcolor[HTML]{FFCCC9}4.92        & \cellcolor[HTML]{FFCCC9}24.09        & \cellcolor[HTML]{9AFF99}(-1.02,-3.27)          &\cellcolor[HTML]{9AFF99} 1.24         & 1.02        & (2.19,-0.11)          & 8.33        & 0.43        \\
		13                        & (-0.70,-1.48)          & 0.68        & 1.21         & (-0.56,-2.42)          & 1.18         & 1.93        & (2.69,-2.88)          & 6.42        & 0.77        \\
		14                        & \cellcolor[HTML]{9AFF99}(0.06,-0.57)           & \cellcolor[HTML]{9AFF99}0.79        & 0.73         & (-1.74,-4.62)          & 1.2          & 1.95        & (-0.79,-0.03)         & 4.43        & 1.78        \\
		15                        & (0.31,-0.26)           & 0.83        & 1.11         & (-1.98,-2.45)          & 2.52         & 3.79        &\cellcolor[HTML]{9AFF99} (-0.56,0.74)          &\cellcolor[HTML]{9AFF99} 5.98        & 0.17        \\
		16                        & (-0.95,-0.48)          & 0.82        & 0.56         & \cellcolor[HTML]{FFCCC9}(-3.64,24.99)          & \cellcolor[HTML]{FFCCC9}3.87         & \cellcolor[HTML]{FFCCC9}33.57       & (-1.58,4.02)          & \cellcolor[HTML]{FFCCC9}1.28        &\cellcolor[HTML]{FFCCC9} 8.48        \\
		17                        & (-1.39,0.67)           & 0.9         & 1.92         & (0.42,-6.47)           & 2.05         & 4.1         & (3.56,2.36)           & 5.35        & 0.83        \\
		18                        & (-1.13,-0.28)          & 0.74        & 0.95         & (0.13,-2.96)           & 1.05         & 2.04        & (3.80,-3.04)          & 6.57        & 0.95        \\
		19                        & (-1.09,-1.00)          & 1.05        & 1.07         & (-0.58,-4.17)          & 1.2          & 1           & (2.91,-1.78)          & 8.26        & 0.67        \\ \hline
	\end{tabular}
	\vspace{-10pt}
\end{table*}

\paragraph{\textbf{Federated Learning Setup}} We utilize the PyTorch package to implement FL in Python. The FL system consists of $N = 60$ clients by default and a central server, which randomly chooses $K = 10$ clients per round for model aggregation. Each training process of FL lasts $T = 200$ rounds. Each training dataset is split into $N$ partitions as the clients' local datasets, which can be either IID or non-IID. The non-IID case is set according to~\cite{marfoq2021federated}.  

% Please add the following required packages to your document preamble:
% \usepackage{multirow}
% \usepackage[table,xcdraw]{xcolor}
% If you use beamer only pass "xcolor=table" option, i.e. \documentclass[xcolor=table]{beamer}

\paragraph{\textbf{Attack Methods}} The four data poisoning attack methods in comparison are set up as follows. 

(1) $Label$ $flipping$ $attack$~\cite{tolpegin2020data}: This method flips specific labels of training data to achieve targeted data poisoning in FL. Specifically, the attacker poisons a local dataset as follows: For all samples belonging to a source class, change their labels to another target class. In our experiment, every data sample with label ``6'' is given a poisoned label ``0''. 

(2) $Label$ $flipping$ $attack$ $with$ $PoisonGAN$~\cite{zhang2020poisongan}: A label flipping attack can work with PoisonGAN, an off-the-shelf GAN model used to enlarge the sizes of local datasets with legitimate pseudo samples by using a discriminator initialized by the global model. Note that PoisonGAN is just a data augmentation method used to enhance the label flipping attack, and it does not directly generate poisoned data. In our experiment, the malicious clients first enlarge their local datasets with pseudo data generated by PoisonGAN and then carry out a label flipping attack as above. The amount of pseudo data generated by PoisonGAN is 10\% of each local dataset. 

(3) $Noise$ $superimposition$ $attack$~\cite{gragnaniello2018analysis}: The attacker superimposes a certain type of noise on real data to generate poisoned data. Either light or heavy Gaussian noise is directly added to the original data, where light noise is with mean = 0 and variance = 0.1; heavy noise is with mean = 0 and variance = 0.3. In addition, SAP noise is added similarly, where 30\% pixels of each image sample are noise-superimposed, and the proportion of ``salt'' versus ``pepper'' noise is 0.5. For cosine noise~\cite{yang2023clean}, a perturbation pattern based on a cosine function is applied to the image data, with an amplitude of 2 and a frequency of 0.5.

(4) $VagueGAN$: For our VagueGAN, it generates the same amount of poisoned data as the real data, and replaces all the real data with the poisoned data. By default, let $\kappa = 0.2$, $E = 600$.

\paragraph{\textbf{Defense Methods}} In response to the above four data poisoning attacks, we compare the following defense methods.

(1) $PCA$~\cite{tolpegin2020data}: This method constructs a local model list on the server, collecting all local models used for federated aggregation. After the local model list is constructed across $T'=80$ rounds, it is standardized by zeroing the mean and being scaled to unit variance. The standardized list is fed into PCA for dimensionality reduction and visualization, and then clients with large outliers are labeled as malicious and subsequently ignored.  

(2) $UMAP$~\cite{upreti2022defending}: This method follows the same logic as PCA, except for using UMAP for dimensionality reduction. As for its settings, neighborhood size for local structure approximation is 200, layout control parameter is 0.8, cosine distance is computed in the ambient space of input data.

(3) $CONTRA$~\cite{awan2021contra}: This method builds upon the PCA-based defense by adding a cosine similarity analysis layer. After performing PCA dimensionality reduction, the cosine similarities between all pairs of clients' gradient vectors are computed. Clients with significantly deviated gradient directions are marked as malicious ones.

(4) $DnC$~\cite{shejwalkar2021manipulating}: This method measures the directional divergence between gradient vectors in PCA space by computing the angles between client updates, identifying poisoned clients through statistically anomalous angular deviations.

(5) $LoMar$~\cite{li2021lomar}: This method applies K-means clustering (with K=2) on PCA-reduced gradient vectors. After dimensionality reduction, it automatically identifies malicious clusters through two criteria: (a) cluster purity (flagging clusters with >80\% poisoned clients) and (b) inter-cluster separation. The neighborhood size for local structure approximation follows the default settings of PCA.

(6) $FedDMC$~\cite{mu2024feddmc}: This method implements a binary tree-based clustering approach on PCA-reduced gradients. The algorithm first constructs a dendrogram using complete linkage hierarchical clustering, and then performs iterative binary splitting based on the maximum intra-cluster distance, while enforcing a minimum cluster size of 2. Potential poisoning clusters are identified when non-majority clusters contain $\geq70\%$ known malicious clients. Distance computations use Euclidean norms in the reduced PCA space.

(7) $MCD$: For our MCD, parameters are set as follows. In MCD, the magnitude of the first part of $h_{f,i}$ in Eq. (21) is slightly larger than that of the second part. To balance the two anomaly evaluation components, we set $\lambda _1=1$ and $\lambda _2=2$. The number of FL rounds for a defense period is $T'=80$. The constraints on the neighborhood for $\mathcal{C}^{\rm{base}}$ are $\varOmega_1$=4, $\varOmega_2$=2.

\begin{figure*}[h]
	\centering
	\includegraphics[width=\linewidth]{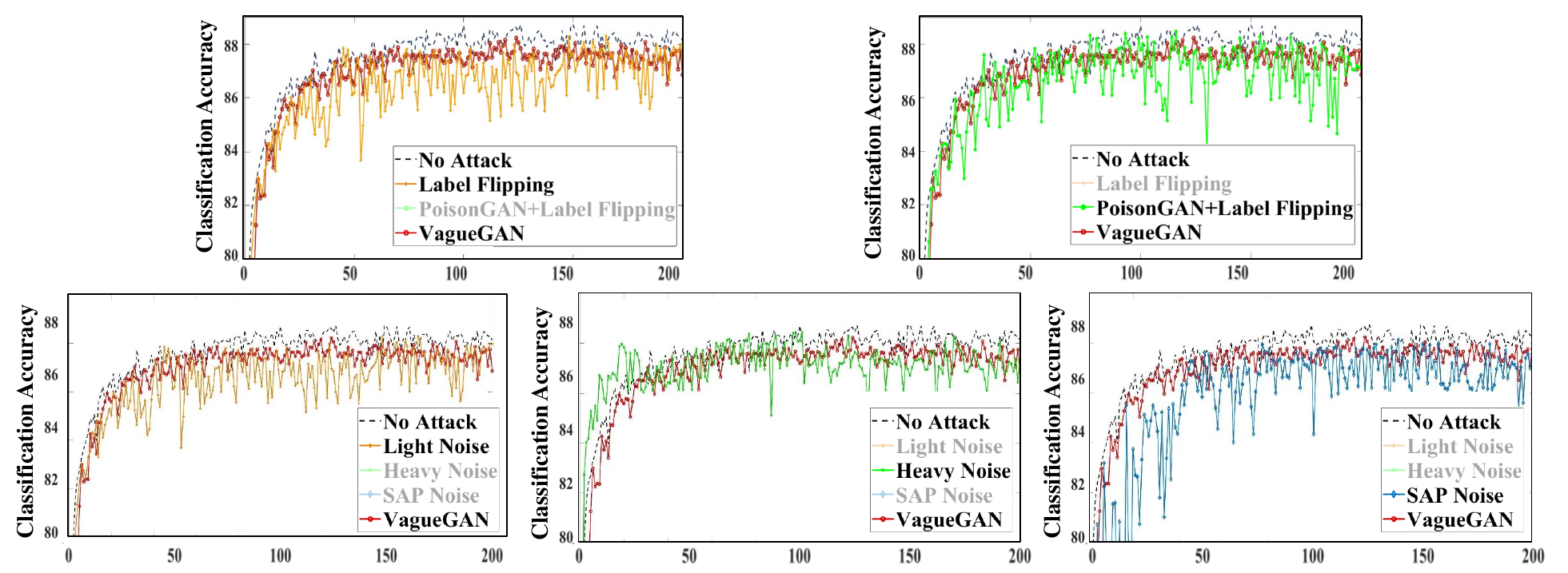}
	\caption{Accuracy fluctuations of attack methods: VagueGAN shows less fluctuations in the global model’s test accuracy than the others, making it harder to detect. }
	\vspace{-15pt}
\end{figure*}

\begin{figure}[h]
	\centering
	\includegraphics[width=\linewidth]{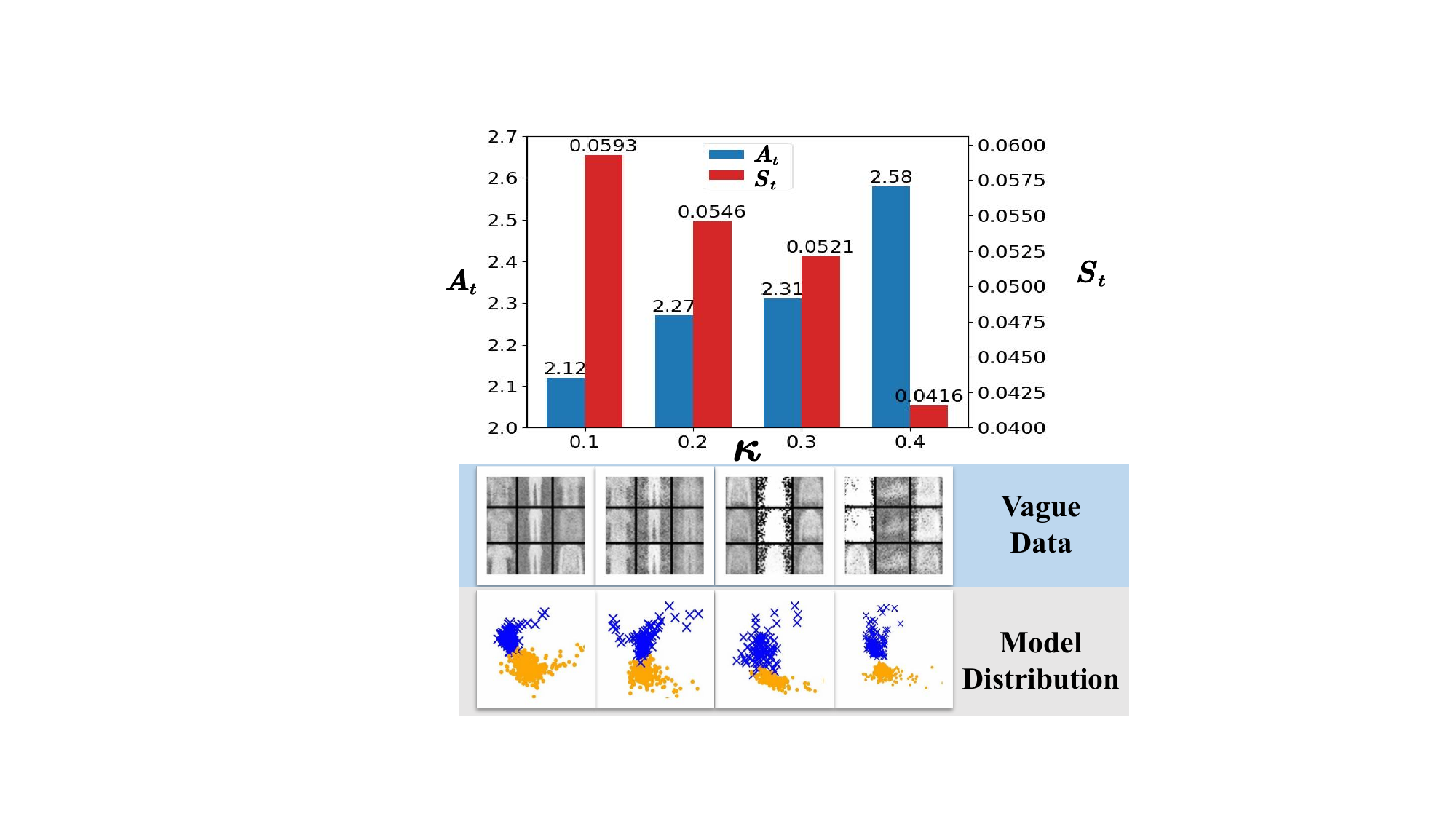}
	\caption{Impact of suppression factor $\kappa$: A larger $\kappa$ gives higher attack effectiveness but lower attack stealthiness.}
\end{figure}
\begin{figure*}[h]
	\centering
	\includegraphics[width=\linewidth]{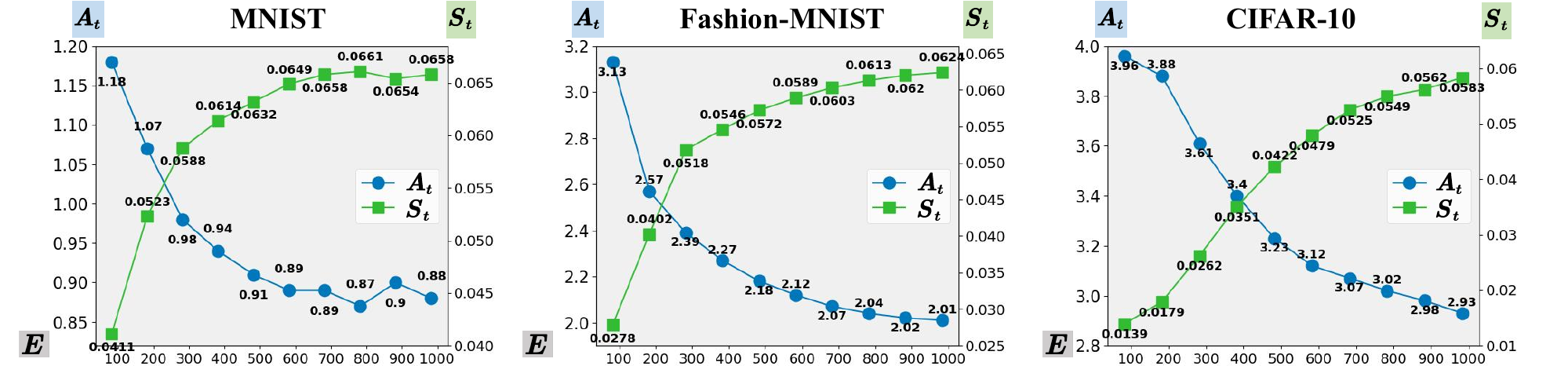}
	\caption{Impact of VagueGAN training epochs $E$: A larger $E$ gives higher attack stealthiness but lower attack effectiveness. }
	\vspace{-8pt}
\end{figure*}

\subsection{Attack Performance Comparison}

%In this subsection, we evaluate and compare the three attack methods in the same scenario. They are implemented to poison the same local datasets as follows. For label flipping, every data sample with label ``6'' is given a poisoned label ``0''. For label flipping with PoisonGAN, the local dataset is first enlarged with pseudo data generated by PoisonGAN and then poisoned by label flipping as above. In ~\cite{zhang2020poisongan}, PoisonGAN generates a small pseudo dataset and inserts it into real data. In our experiment, the amount of pseudo data generated by PoisonGAN is 10\% of each local dataset. For noise superimposition, either light or heavy Gaussian noise is directly added to the original data, where light noise is with mean = 0 and noise variance = 0.1; heavy noise is with mean = 0 and noise variance = 0.3. For VagueGAN, it generates the same amount of poisoned data as the real data, and replace the real data with the poisoned data, and let $\kappa = 0.2$, $E = 600$.

In this subsection, we evaluate and compare the four attack methods in the same scenario, all of which are implemented to poison the same local datasets. It is worth noting that, to ensure a fair comparison, all the attack methods including VagueGAN use the same hyperparameters for the effectiveness and stealthiness evaluation experiments in subsubsections VI-B-a and VI-B-b, respectively, namely $\kappa = 0.2$ and $E = 600$. In other words,  the effectiveness of VagueGAN and in subsubsection VI-B-a and the stealthiness of it in subsubsection VI-B-b are achieved under the same experimental settings.

\paragraph{\textbf{Effectiveness of Poisoning Attacks}} It is evident that the most intuitive measure of the effectiveness of a data poisoning attack is the degree to which the test performance (i.e., task accuracy) of the global model on the main task degrades following the attack. Consequently, the attack effectiveness of data poisoning in this context is characterized by the global model accuracy $a^p_t$ when the attack is present, compared to the no-attack accuracy $a_t$ under the same conditions. For more details, refer to subection IV-D. With the percentage of malicious clients $\alpha$ ranged from 5\% to 30\%, the effectiveness of the four attack methods is compared in Table I, and the results are averaged after 10 runs. Compared with the benchmark methods, our VagueGAN mostly achieves better attack effectiveness, though its major advantage is the trade-off between attack effectiveness and stealthiness. Not surprisingly, increasing the percentage of malicious clients and a non-IID setting lower the global model accuracy. In addition, the attack methods perform differently on different datasets. Compared with Fashion-MNIST and MNIST, CIFAR-10 and CIFAR-100, as well as Mini-ImageNet, are more vulnerable to data poisoning attacks. To explain, we name the rounds in which the number of malicious clients selected by the server is greater than $2$ as abnormal rounds, and the rest of the rounds as normal rounds. We find that after any abnormal round, the accuracy rate of the global model generally first decreases, and then returns to normal after a few normal rounds. In comparison, more such normal rounds are needed for CIFAR-10, CIFAR-100, and Mini-ImageNet than for Fashion-MNIST and MNIST. This is because these datasets make the FL task more complex, thereby requiring more normal rounds to recover.

%Additionally, we evaluate the effectiveness of the intuitive noise superimposition attack. Take the Fashion-MNIST dataset under IID as an example, and let $\alpha=20\%$. As shown in Figure 8, the two noise lines correspond to the data in Figure 4 with Gaussian noise added, where light noise is with noise mean = 0, noise variance = 0.1; heavy noise is with noise mean = 0, noise variance = 0.3. The test accuracy values of the no noise, light noise, heavy noise and VagueGAN cases are fitted to 88.12\%, 87.97\%, 87.16\% and 86.66\%, respectively. Obviously, the noise level of the light noise case is not enough to achieve an effective attack. Although the noise level of the heavy noise case can be increased to carry out an attack as effective as that achieved by VagueGAN, later we will show that the noise superimposition is always worse than our VagucGAN in terms of attack stealthiness. 

\paragraph{\textbf{Stealthiness of Poisoning Attacks}} In this paper, we characterize the stealthiness of a data poisoning attack using statistical distances, such as those obtained between legitimate and poisoned models after applying PCA. This approach is adopted because most existing defenses against data poisoning follow a similar strategy~\cite{tolpegin2020data,upreti2022defending,shejwalkar2021manipulating,zhang2022fldetector,shen2023privacy,awan2021contra,li2021lomar}. They typically obtain low -dimensional model parameters to facilitate subsequent analysis, and use Euclidean distance, cosine similarity, or clustering to measure model differences. This approach can indirectly identify poisoned models without the need for accessing the original data. Take the Fashion-MNIST task under the setting of IID and $\alpha=20\%$ as an example. After compressing local models to 2-dimensional for visualization using PCA, as shown in Figure 9, the statistical distance between legitimate models (yellow Os for the last $50$ rounds) and poisoned models (blue Xs) can be measured. Similarly, the local models can also be visualized in higher dimensions, and our VagueGAN consistently achieves much better attack stealthiness than the benchmark methods. 
%First, the attack stealthiness of data poisoning is characterized by the statistical distances between legitimate and poisoned models obtained after applying e.g., PCA. Most of the existing defense methods follow a similar idea.         

%\begin{figure}[h]
%	\centering
%	\includegraphics[width=\linewidth]{noise yinbi}
%	\caption{Local model visualization under noise superimposition attack}
%\end{figure}

\begin{table*}[htbp]
	\caption{Defense effectiveness of MCD in different attack scenarios with varying proportions of malicious clients.}
	\begin{center}%{width=\textwidth}
		\setlength{\tabcolsep}{2.78mm}
		\begin{tabular}{c|c|c|c|c|c|c|c|c}
			\hline
			&
			\textbf{M} &
			$\boldsymbol{\alpha }$ &
			\textbf{Label flipping} &
			\textbf{PoisonGAN} &
			\textbf{Light noise} &
			\textbf{Heavy noise} &
			\textbf{SAP noise} &
			\textbf{VagueGAN} \\ \hline
			\multirow{4}{*}{\begin{tabular}[c]{@{}c@{}}Number of \\ malicious clients \\ detected by \\ MCD (average)\end{tabular}} &
			1 &
			5\% &
			1 &
			1 &
			1 &
			1 &
			1 &
			1 \\  
			&
			2 &
			10\% &
			2 &
			2 &
			1.9 &
			2 &
			2 &
			2 \\ 
			&
			4 &
			20\% &
			4 &
			4 &
			3.8 &
			4 &
			3.9 &
			3.9 \\ 
			&
			6 &
			30\% &
			6 &
			6 &
			5.6 &
			6 &
			5.7 &
			5.8 \\ \hline
			&
			$\boldsymbol{a_t }$ &
			$\boldsymbol{\alpha }$ &
			\textbf{Label flipping} &
			\textbf{PoisonGAN} &
			\textbf{Light noise} &
			\textbf{Heavy noise} &
			\textbf{SAP noise} &
			\textbf{VagueGAN} \\ \hline
			\multirow{4}{*}{\begin{tabular}[c]{@{}c@{}}Main task \\ test accuracy after \\MCD deployment\\ (average)(\%)\end{tabular}} &
			\multirow{4}{*}{87.97} &
			5\% &
			87.94 ($\uparrow$0.21) &
			87.92 ($\uparrow$0.33) &
			87.97 ($\uparrow$0.16) &
			87.95 ($\uparrow$0.24) &
			87.95 ($\uparrow$0.15) &
			87.94 ($\uparrow$0.62) \\  
			&
			&
			10\% &
			87.91 ($\uparrow$0.71) &
			87.92 ($\uparrow$0.75) &
			87.90 ($\uparrow$0.32) &
			87.87 ($\uparrow$0.72) &
			87.89 ($\uparrow$0.37) &
			87.89 ($\uparrow$1.11) \\ 
			&
			&
			20\% &
			87.88 ($\uparrow$1.87) &
			87.84 ($\uparrow$2.07) &
			87.87 ($\uparrow$0.95) &
			87.83 ($\uparrow$1.77) &
			87.85 ($\uparrow$0.86) &
			87.86 ($\uparrow$1.99) \\ 
			&
			&
			30\% &
			87.84 ($\uparrow$2.36) &
			87.80 ($\uparrow$2.71) &
			87.86 ($\uparrow$1.49) &
			87.82 ($\uparrow$2.26) &
			87.90 ($\uparrow$1.47) &
			87.89 ($\uparrow$3.12) \\ \hline
		\end{tabular}
	\end{center}
	\vspace{-8pt}
\end{table*}

\begin{table*}[htbp]
	\caption{Defense effectiveness of PCA in different attack scenarios with varying proportions of malicious clients.}
	\begin{center}%{width=\textwidth}
		\setlength{\tabcolsep}{2.78mm}
		\begin{tabular}{c|c|c|c|c|c|c|c|c}
			\hline
			&
			\textbf{M} &
			$\boldsymbol{\alpha }$ &
			\textbf{Label flipping} &
			\textbf{PoisonGAN} &
			\textbf{Light noise} &
			\textbf{Heavy noise} &
			\textbf{SAP noise} &
			\textbf{VagueGAN} \\ \hline
			\multirow{4}{*}{\begin{tabular}[c]{@{}c@{}}Number of \\ malicious clients \\ detected by \\ PCA (average)\end{tabular}} &
			1 &
			5\% &
			0.9 &
			1 &
			0.6 &
			0.8 &
			0.8 &
			0.1 \\  
			&
			2 &
			10\% &
			1.7 &
			1.9 &
			1.4 &
			1.7 &
			1.8 &
			0.2 \\ 
			&
			4 &
			20\% &
			3.5 &
			3.6 &
			2.7 &
			3.5 &
			3.6 &
			0.6 \\ 
			&
			6 &
			30\% &
			5.3 &
			5.5 &
			4.2 &
			5.1 &
			5.3 &
			0.9 \\ \hline
			&
			$\boldsymbol{a_t }$ &
			$\boldsymbol{\alpha }$ &
			\textbf{Label flipping} &
			\textbf{PoisonGAN} &
			\textbf{Light noise} &
			\textbf{Heavy noise} &
			\textbf{SAP noise} &
			\textbf{VagueGAN} \\ \hline
			\multirow{4}{*}{\begin{tabular}[c]{@{}c@{}}Main task \\ test accuracy after \\PCA deployment\\ (average)(\%)\end{tabular}} &
			\multirow{4}{*}{87.97} &
			5\% &
			87.92 ($\uparrow$0.19) &
			87.89 ($\uparrow$0.30) &
			87.93 ($\uparrow$0.12) &
			87.92 ($\uparrow$0.21) &
			87.92 ($\uparrow$0.12) &
			87.34 ($\uparrow$0) \\  
			&
			&
			10\% &
			87.86 ($\uparrow$0.66) &
			87.87 ($\uparrow$0.67) &
			87.79 ($\uparrow$0.21) &
			87.76 ($\uparrow$0.61) &
			87.78 ($\uparrow$0.26) &
			86.79 ($\uparrow$0) \\ 
			&
			&
			20\% &
			87.79 ($\uparrow$1.78) &
			87.77 ($\uparrow$1.76) &
			87.71 ($\uparrow$0.79) &
			87.56 ($\uparrow$1.50) &
			87.78 ($\uparrow$0.79) &
			86.01 ($\uparrow$0.14) \\ 
			&
			&
			30\% &
			87.66 ($\uparrow$2.18) &
			87.61 ($\uparrow$2.52) &
			87.68 ($\uparrow$1.31) &
			87.49 ($\uparrow$1.93) &
			87.82 ($\uparrow$1.39) &
			84.97 ($\uparrow$0.20) \\ \hline
		\end{tabular}
	\end{center}
	\vspace{-8pt}
\end{table*}

\begin{table*}[htbp]
	\caption{Defense effectiveness of UMAP in different attack scenarios with varying proportions of malicious clients.}
	\begin{center}%{width=\textwidth}
		\setlength{\tabcolsep}{2.78mm}
		\begin{tabular}{c|c|c|c|c|c|c|c|c}
			\hline
			&
			\textbf{M} &
			$\boldsymbol{\alpha }$ &
			\textbf{Label flipping} &
			\textbf{PoisonGAN} &
			\textbf{Light noise} &
			\textbf{Heavy noise} &
			\textbf{SAP noise} &
			\textbf{VagueGAN} \\ \hline
			\multirow{4}{*}{\begin{tabular}[c]{@{}c@{}}Number of \\ malicious clients \\ detected by \\ UMAP (average)\end{tabular}} &
			1 &
			5\% &
			0.7 &
			0.8 &
			0.4 &
			0.8 &
			0.8 &
			0 \\  
			&
			2 &
			10\% &
			1.4 &
			1.9 &
			1.2 &
			1.7 &
			1.5 &
			0 \\ 
			&
			4 &
			20\% &
			2.9 &
			2.8 &
			1.9 &
			2.3 &
			2.7 &
			0.2 \\ 
			&
			6 &
			30\% &
			3.3 &
			4.0 &
			2.9 &
			3.7 &
			4.1 &
			0.7 \\ \hline
			&
			$\boldsymbol{a_t }$ &
			$\boldsymbol{\alpha }$ &
			\textbf{Label flipping} &
			\textbf{PoisonGAN} &
			\textbf{Light noise} &
			\textbf{Heavy noise} &
			\textbf{SAP noise} &
			\textbf{VagueGAN} \\ \hline
			\multirow{4}{*}{\begin{tabular}[c]{@{}c@{}}Main task \\ test accuracy after \\UMAP deployment\\ (average)(\%)\end{tabular}} &
			\multirow{4}{*}{87.97} &
			5\% &
			87.92 ($\uparrow$0.19) &
			87.87 ($\uparrow$0.28) &
			87.93 ($\uparrow$0.12) &
			87.91 ($\uparrow$0.20) &
			87.94 ($\uparrow$0.14) &
			87.34 ($\uparrow$0) \\  
			&
			&
			10\% &
			87.82 ($\uparrow$0.62) &
			87.89 ($\uparrow$0.69) &
			87.79 ($\uparrow$0.21) &
			87.72 ($\uparrow$0.57) &
			87.70 ($\uparrow$0.18) &
			86.79 ($\uparrow$0) \\ 
			&
			&
			20\% &
			87.57 ($\uparrow$1.56) &
			87.61 ($\uparrow$1.60) &
			87.59 ($\uparrow$0.67) &
			87.43 ($\uparrow$1.37) &
			87.62 ($\uparrow$0.63) &
			85.96 ($\uparrow$0.09) \\ 
			&
			&
			30\% &
			87.34 ($\uparrow$1.86) &
			87.39 ($\uparrow$2.30) &
			87.61 ($\uparrow$1.24) &
			87.29 ($\uparrow$1.73) &
			87.60 ($\uparrow$1.17) &
			84.91 ($\uparrow$0.14) \\ \hline
		\end{tabular}
	\end{center}
	\vspace{-15pt}
\end{table*}

Second, the attack stealthiness of data poisoning can also be characterized by fluctuations in the global model accuracy $a^p_t$. Such accuracy fluctuations are linked to attack stealthiness because the server's random selection of clients per round may bring more or less possibly noticeable poisoned models and thus may cause the accuracy fluctuations. Taking the Fashion-MNIST task under the setting of IID and $\alpha=20\%$ as an example, the accuracy level fluctuating with FL rounds given by VagueGAN versus other data poisoning attack methods is shown in Figure 10. Clearly, our VagueGAN with small accuracy fluctuations mostly approximates the no attack case, but the benchmark methods result in relatively large accuracy fluctuations and are easily detectable even without statistical analysis. Therefore, higher stealthiness of VagueGAN poisoning attack makes it easier to be successful. 

%Meanwhile, we also evaluate the stealthiness of the noise superimposition attack set as above. First, we visualize 2-dimensional local models under different noise levels in Figure 10. It can be seen that whether there is light noise or heavy noise, the stealthiness of the noise superimposition is not very good. This is because the distinct distribution of noise greatly affects the distribution of real data, resulting in large distances between legitimate and poisoned models. Second, it can be seen from Figure 7 that the accuracy of the global model fluctuates noticeably under a heavy noise level. In summary, when the noise level of noise superimposition attack is low, both the effectiveness and stealthiness are not ideal; when the noise level is high, the low stealthiness prevents this method from being useful. 

%\begin{figure}[h]
%	\centering
%	\includegraphics[width=\linewidth]{all4}
%	\caption{VagueGAN versus label flipping accuracy fluctuations}
%\end{figure}

\subsection{Performance Trade-off Analysis}

As discussed in subsection IV-C, a balanced trade-off between attack effectiveness $A_t$ and stealthiness $S_t$ can be achieved by setting the values of $\kappa$ and $E$ appropriately. 

\paragraph{\textbf{Generator Suppression Factor $\kappa$}} The value of $\kappa$ can be adjusted to limit the generation ability of the generator of VagueGAN. Taking the Fashion-MNIST task under the setting of IID, $\alpha=20\%$, and $E=400$ as an example, the impact of $\kappa$ on the trade-off between $A_t$ and $S_t$ is evaluated in Figure 11. It can be seen that increasing $\kappa$ from 0.1 to 0.4 gives increasing $A_t$ but decreasing $S_t$. To ensure a not easily detectable attack, a reasonable value of $\kappa$, e.g., 0.2, should not be too large.   

\begin{figure}[h]
	\centering
	\includegraphics[width=\linewidth]{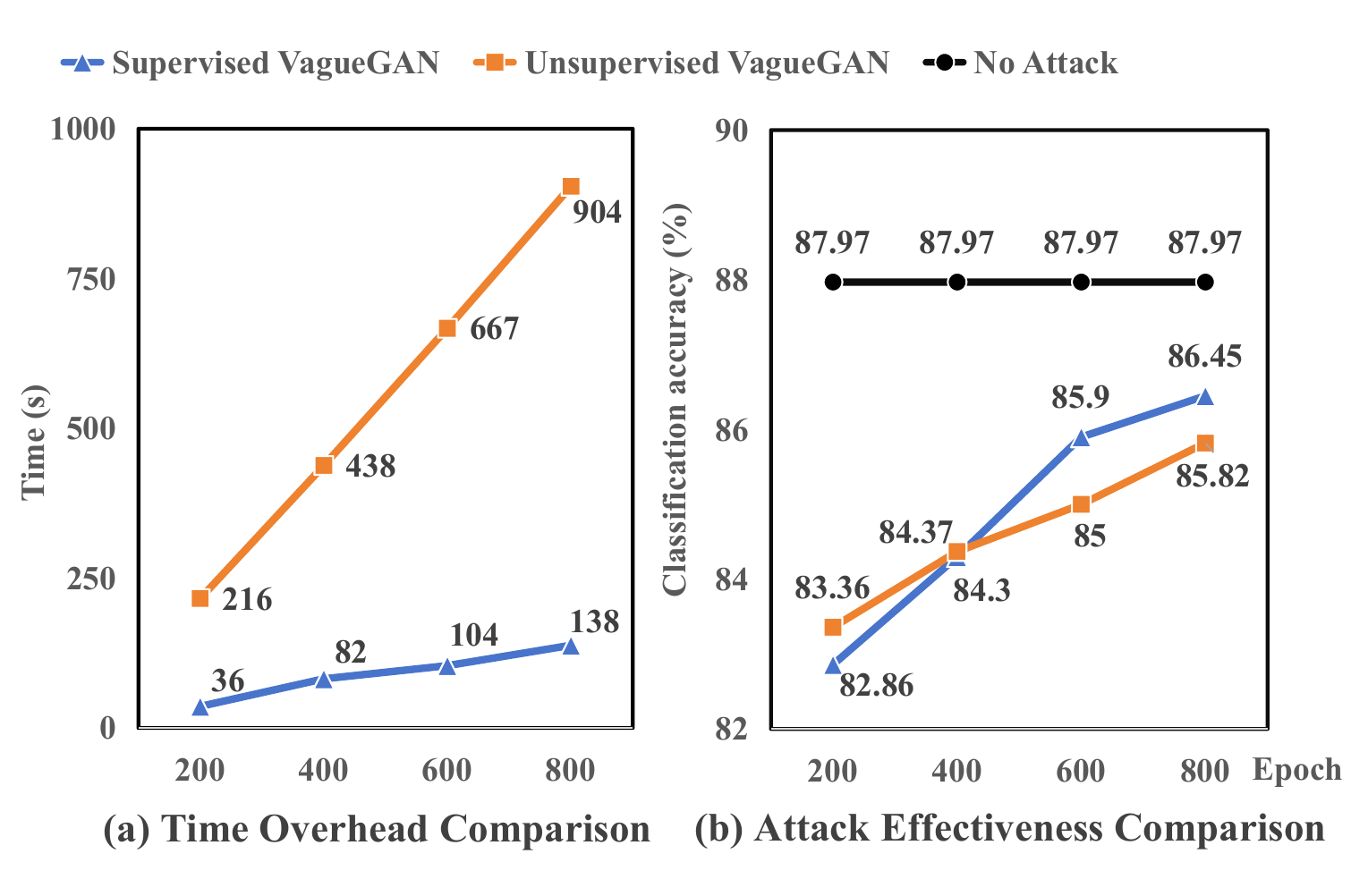}
	\caption{Supervised vs. unsupervised VagueGAN for time overhead and attack effectiveness: Unsupervised VagueGAN requires higher time overhead but achieves less fluctuant and comparable attacks without label information. }
	\vspace{-5pt}
\end{figure}

\paragraph{\textbf{Number of Training Epochs $E$}} The value of $E$ can be set to control the fitting state of VagueGAN. When $\alpha=20\%$, the impact of $E$ on the trade-off between $A_t$ and $S_t$ is evaluated in Figure 12. It can be seen that with the increase of $E$, effectiveness $A_t$ gradually decreases, while stealthiness $S_t$ gradually increases. The rate of change for both $A_t$ and $S_t$ decreases with the increase of $E$. On the one hand, the value of $E$ cannot be too small. We find that according to the visualized model distribution, when $S_t<0.05$, the corresponding data poisoning attack will become detectable. Although the performance of $A_t$ is satisfactory at this time, the attack cannot be successful if given a regular defense system. On the other hand, the value of $E$ cannot be too large. For the Fashion-MNIST task as an example, we find that the vague data generated by VagueGAN will be no longer ``vague'' when $E>600$ and the quality of vague data will no longer change when $E>1000$. For this case, the performance of $A_t$ can be overly harmed. More generally, we find that an appropriate value of $E$ can be $0.5e_{lim}$, where $e_{lim}$ is the number of training epochs required for VagueGAN to reach a converged steady state, e.g., $e_{lim}$= 600 for MNIST, 1000 for Fashion-MNIST, and 1600 for CIFAR-10 and CIFAR-100, , and 2000 for Mini-ImageNet.

\subsection{Comparison of supervised and unsupervised VagueGANs}

In this subsection, we present an experimental comparison of supervised VagueGAN with its unsupervised variant. We evaluate their time overhead, attack effectiveness, and stealthiness with the same settings as follows: generator suppression factor $\kappa$=0.2; the number of training epochs $E$ are taken as 200, 400, 600, and 800, respectively; and the same amount of poisoned data is generated. Firstly, the time overheads are compared for the two VagueGANs on the same experimental device, with a GPU:RTX 3090(24GB) and a CPU:12 vCPU Intel(R) Xeon(R) Platinum 8255C CPU. Figure 13(a) illustrates that the time overhead of unsupervised VagueGAN is greater than that of supervised VaugeGAN. Furthermore, the larger the number of training epochs, the greater the discrepancy between the two can be observed. This is due to the necessity that unsupervised VagueGAN maintains an additional classifier, $Q$, and learns the auxiliary features of the training data.
\begin{figure}[h]
	\centering
	\includegraphics[width=\linewidth]{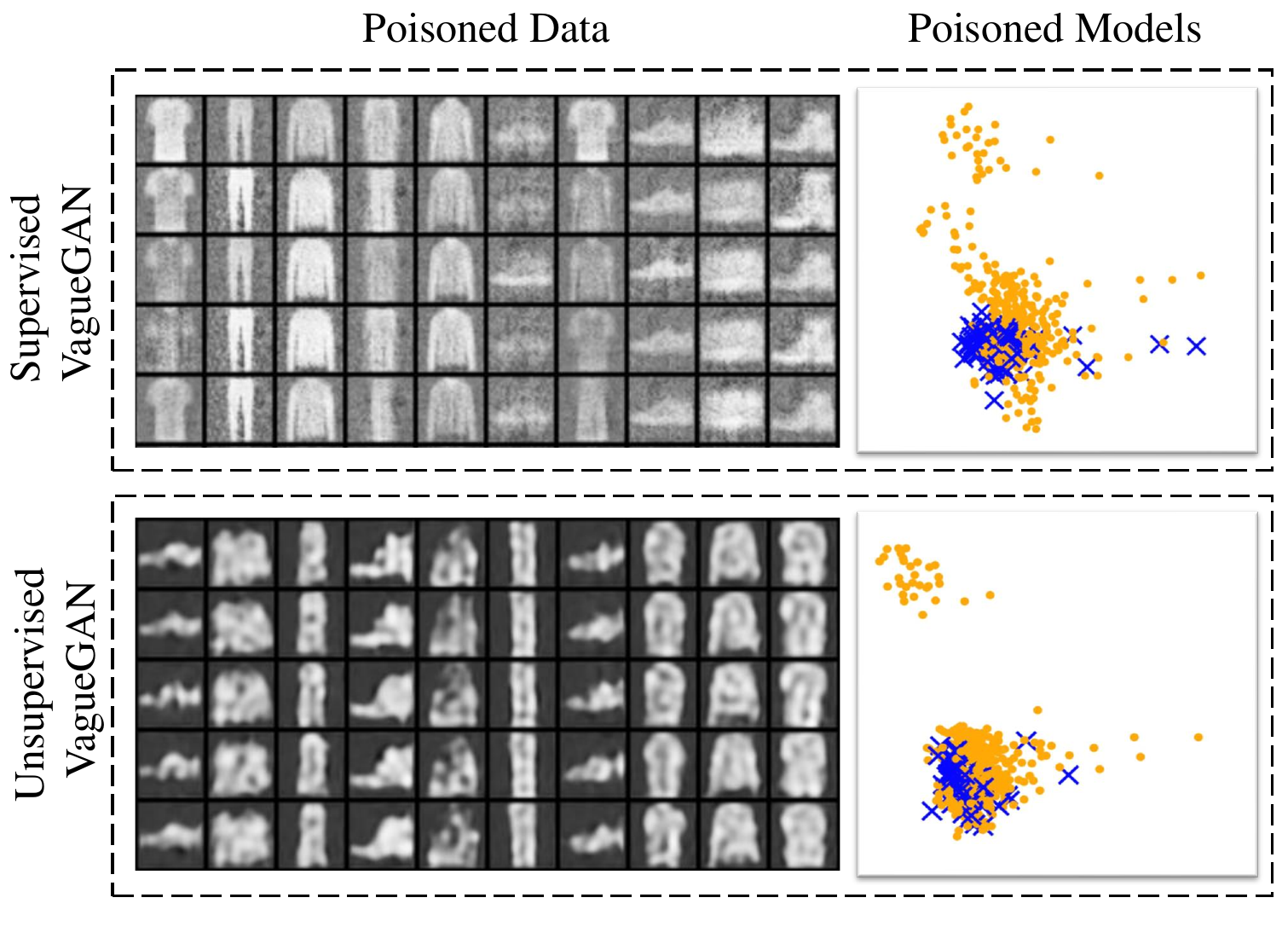}
	\caption{Supervised vs. unsupervised VagueGAN for attack stealthiness: Both achieve high stealthiness, and poisoned models can look more normal than some legitimate ones. }
\end{figure}
\begin{figure*}[h]
	\centering
	\includegraphics[width=\linewidth]{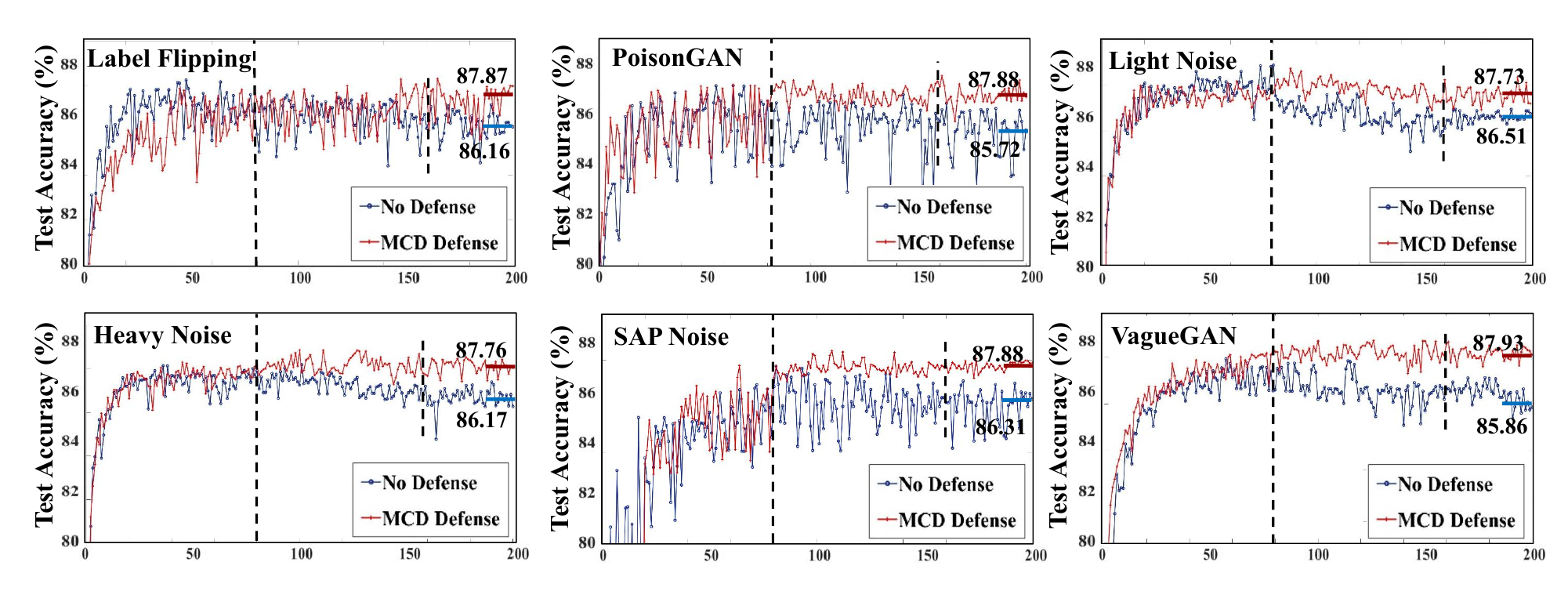}
	\caption{Defense effectiveness of MCD against different attack methods: Accuracy improvements can be observed when MCD is applied.}
	\vspace{-15pt}
\end{figure*}

Secondly, we compare the attack effectiveness of two VagueGANs. Figure 13(b) demonstrates that both VagueGANs exhibit excellent, comparable attack effectiveness. It is noteworthy that the unsupervised attack is more stable and less susceptible to fluctuations in training epochs.

Finally, we compare the attack stealthiness of the two VagueGANs. As illustrated in Figure 14, the outcomes are compared under the trade-off between effectiveness and stealthiness ($\kappa=0.2, E=600$), indicating excellent stealthiness. Furthermore, it is evident that both VagueGANs not only harm the global model performance but also indirectly mislead the distribution of some benign client models and make them look like outliers.

\subsection{Defense Performance Comparison}

In this subsection, we evaluate the effectiveness of MCD against different data poisoning attacks in FL, including label flipping attack, label flipping attack with PoisonGAN, noise superimposition attack and VagueGAN. The FL system is set as follows: the total number of clients $N=20$; the number of clients selected by the server in each round $K=5$; the number of FL training rounds $T=200$; the main task dataset is Fashion-MNIST under the setting of IID. We experiment a case where the percentage of malicious clients $\alpha$ is 20\%.

\paragraph{\textbf{Anomaly Detection Capability of MCD}} Tables II and III are the metric tables of MCD in different attack scenarios. Among them, $i$ denotes the client identifier, where red numbers represent malicious clients and green numbers represent the primary benchmark client. $\boldsymbol{\bar{\theta}}_{f,i}^{(2)}$ and $\bar{d}_{f,i}$ are the center and footprint metrics, respectively, used by the MCD for anomaly detection. $\boldsymbol{{\theta}}^{\rm{base}}_f$ and ${d}_{f}^{\rm{base}}$ serve as the benchmark metrics calculated by the MCD. Light green indicators represent the 100\% safe benchmarks $(\boldsymbol{\hat{\theta}}_{f}^{(2)}, \hat{d}_{f})$, while light red indicators denote the anomalies detected by the MCD. The anomaly degree for each client, $h_{f,i}$, is computed by the MCD, with red indicating anomalous clients. For detailed definitions of these metrics, please refer to subection V-B.

Table II present the results of the label flipping attacks and label flipping attacks with PoisonGAN, and ight noise superimposition attack. It is evident from the tables that the $\boldsymbol{\bar{\theta}}_{f,i}^{(2)}$ of the malicious clients exhibit pronounced anomalies in both attack scenarios, and the abnormal behavior of the light noise superimposition attack is similar to that of label flipping attack, as both exhibit anomalies in $\boldsymbol{\bar{\theta}}_{f,i}^{(2)}$. In the scenario of heavy noise and SAP noise superimposition attack as shown in Tables III, the malicious clients demonstrate anomalies in both $\boldsymbol{\bar{\theta}}_{f,i}^{(2)}$ and $\bar{d}_{f,i}$, indicating a poor level of stealthiness.

%Upon analysis, it can be observed that the model distributions of malicious clients deviate significantly in these two attack scenarios. Furthermore, the malicious client abnormality $h_{t,i}$ is noticeably higher than those of benign clients. Hence, MCD successfully and accurately identifies these malicious clients.

Table III displays the judgment results of MCD in the data poisoning attack scenario based on VagueGAN. Unlike the previous attacks, the malicious clients in the VagueGAN attack exhibit no anomalies in $\boldsymbol{\bar{\theta}}_{f,i}^{(2)}$. This indicates that the malicious clients taking advantage of GAN have largely learned the distribution of legitimate models and thus are hard to detect if using traditional defense methods like PCA. However, the malicious clients exhibit significantly lower $\bar{d}_{f,i}$ values compared with the normal values of benign clients, meaning that MCD still works for GAN-based attacks like VagueGAN.

% Please add the following required packages to your document preamble:
% \usepackage{multirow}

\paragraph{\textbf{Overall Defense Effectiveness of MCD}}

%In order to better demonstrate the ability of MCD to identify abnormal parameters, we visualize the above parameters.

%As shown in Figure 13, the metrics $\boldsymbol{\bar{\theta}}_{t,i}^{(2)}$ of the label flipping attack model show a very serious abnormality. The corresponding metrics of the data poisoning attack based on VagueGAN have no abnormality, because it has taken corresponding concealment measures to disguise its malicious model. As shown in Figure 14, the data poisoning attack based on PoisonGAN has abnormalities in $\bar{d}_{t,i}$, which is represents the shape of the model distribution. As we analyzed in subsection 5.1, the distribution of malicious client models for data poisoning attacks using VagueGAN is unusually dense, and MCD is well positioned to detect anomalies using this feature. Figure 15 shows the results of the visualization of the anomaly degrees $h_{t,i}$, it can be seen that MCD has a good detection effect on these four iconic data poisoning attacks.

%In addition, we conduct experiments in four attack scenarios and with MCD deployed respectively. Figure 15 shows the improvement effect of FL master task training accuracy, respectively. The dotted line is the timing for MCD to run the anomaly detection algorithm. In all attack scenarios, MCD identified all abnormal clients during the first run. The anomaly detection algorithm is then run every 40 rounds to ensure the safety of the FL system. It can be seen that after the deployment of MCD, the accuracy of the main task in all attack scenarios has increased. 
Figure 15 presents the gain in FL performance brought by MCD in six distinct attack scenarios. It is evident that the incorporation of MCD significantly enhances the accuracy of the main task within the FL system. Combining Tables III and IV to Figure 15 and specific experimental observations, we found across all the six attack scenarios, MCD successfully identifies malicious clients during the first inspection and disregards their subsequent updates (as indicated by the long dashed line). To ensure a stable security level for the FL system, MCD performs regular executions in subsequent training rounds (as indicated by the short dashed line).

We further evaluate MCD in attack scenarios with different proportions of malicious clients, as shown in Table IV. The table presents results, on the upper half, for the number of malicious clients detected by MCD, and on the lower half, for the accuracy improvement of the global model brought by MCD. Herein, $M$ (or $\alpha$)denotes the number (or percentage) of malicious clients within the FL system, and $a_t$ denotes the global model accuracy in the absence of malicious clients. Each scenario is performed 10 times for average results. It is shown that MCD can always identify most of the malicious clients as in the upper half table, and thus achieve a gain in FL performance as in the lower half table. 

We also compare MCD with PCA and UMAP, two typical existing approaches, as shown in Tables V and VI respectively. The results are produced from the same attack scenarios as in Table IV. Each scenario is also performed 10 times for average results. It can be seen that MCD always outperforms PCA and UMAP in terms of the number of malicious clients identified and accuracy improvement, especially under the VagueGAN. 

Furthermore, the detection results of four additional defense methods against VagueGAN are evaluated. In Figures 16(a) and 16(b), the cosine similarity results given by CONTRA and the gradient vector angle deviation results given by DnC are presented for the models of three types of client pairs: benign-benign, poisoned-poisoned and benign-poisoned client pairs. 
The results indicate that the models of the poisoned clients exhibit no significant difference in the deviation statistics of both cosine similarity and vector angle difference distributions do not significantly differ from those of benign clients, indicating that neither CONTRA nor DnC can successfully detect stealthier attacks like VagueGAN. Meanwhile, Figures 16(c) and 16(d) display the results obtained using different clustering methods LoMar and FedDMC, with distinct clusters represented by different colors and malicious clients marked with crosses. Because the feature distribution of poisoned client models is somewhat different from that of benign client models, ideally they can be clustered into different clusters to tell them apart. It can be seen however the clustering results from both LoMar and FedDMC fail to effectively distinguish the malicious and benign client models, indicating that neither LoMar nor FedDMC can successfully detect stealthier attacks like VagueGAN. Therefore, our MCD outperforms these four defense methods in working against GAN-based data poisoning attacks with enhanced attack stealthiness. 

\begin{figure}[h]
	\centering
	\includegraphics[width=\linewidth]{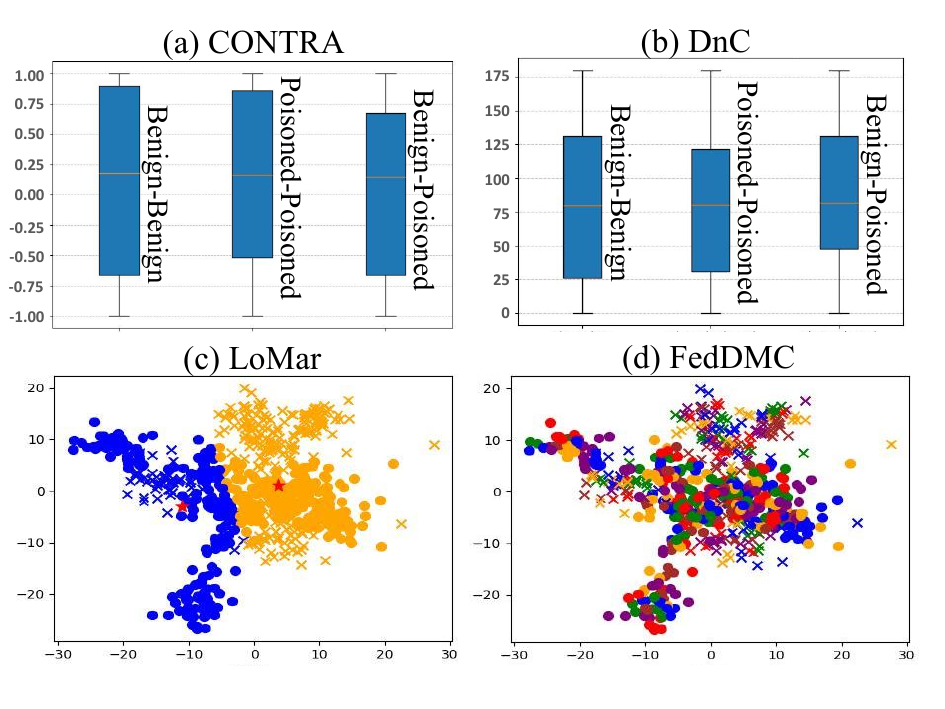}
	\caption{Detection results of different defenses against VagueGAN: None of them achieves successful detection. }
	\vspace{-5pt}
\end{figure}
%These experimental findings collectively attest to the efficacy of MCD as a robust defense mechanism against diverse data poisoning attacks, characterized by its precision and efficiency in the defense processes.

As summarized in Table VII, our MCD can effectively work against various data poisoning attacks, especially GAN-based ones like our VagueGAN, and can further contribute to the recovery of FL performance under attacks. It is important to note that this study primarily aims to investigate the impact of poisoned data on FL, so we focus on the case where poisoned data is generated by a consistently applied VagueGAN model. Nonetheless, if the attacker were able to adopt a more dynamic version of VagueGAN or integrate it with more sophisticated adaptive attack strategies, the robustness of MCD would be subject to significantly greater challenges.

\begin{table}[ht]
	\caption{Defense methods vs. attack methods: A check mark (or a cross mark) indicates that the defense can (or cannot) counter the corresponding attack.}
	\label{tab:my-table}
	\begin{tabular}{c|>{\centering\arraybackslash}p{0.62cm} 
			>{\centering\arraybackslash}p{0.62cm}
			>{\centering\arraybackslash}p{0.62cm}
			>{\centering\arraybackslash}p{0.62cm}
			>{\centering\arraybackslash}p{0.62cm}
			>{\centering\arraybackslash}p{0.62cm}
			>{\centering\arraybackslash}p{0.62cm}} % Adjust the column width
		\hline
		& \multicolumn{7}{c}{Attack} \\ \cline{2-8} 
		\multirow{-2}{*}{Defense} & \begin{tabular}[c]{@{}c@{}}Label \\ Flipping\end{tabular} & \begin{tabular}[c]{@{}c@{}}Poison\\ GAN\end{tabular} & \begin{tabular}[c]{@{}c@{}}Light \\ Noise\end{tabular} & \begin{tabular}[c]{@{}c@{}}Heavy \\ Noise\end{tabular} & \begin{tabular}[c]{@{}c@{}}SAP \\ Noise\end{tabular} & \begin{tabular}[c]{@{}c@{}}Cosine \\ Noise\end{tabular} & \begin{tabular}[c]{@{}c@{}}\textbf{Vague} \\ \textbf{GAN}\end{tabular} \\ \hline
		PCA                       & {  \checkmark}                                  & {  \checkmark}                             & {   \texttimes}                               & {  \checkmark}                               & {   \texttimes}                             & {   \texttimes}                                & {   \texttimes}                             \\
		UMAP                      & {  \checkmark}                                  & {  \checkmark}                             & {   \texttimes}                               & {   \texttimes}                               & {   \texttimes}                             & {   \texttimes}                                & {   \texttimes}                             \\
		CONTRA                    & {  \checkmark}                                  & {  \checkmark}                             & {   \texttimes}                               & {  \checkmark}                               & {   \texttimes}                             & {   \texttimes}                                & {   \texttimes}                             \\
		DnC                       & {  \checkmark}                                  & {  \checkmark}                             & {   \texttimes}                               & {  \checkmark}                               & {  \checkmark}                             & {  \checkmark}                                & {   \texttimes}                             \\
		Lomar                     & {  \checkmark}                                  & {  \checkmark}                             & {   \texttimes}                               & {   \texttimes}                               & {   \texttimes}                             & {   \texttimes}                                & {   \texttimes}                             \\
		FedDMC                    & {  \checkmark}                                  & {  \checkmark}                             & {   \texttimes}                               & {  \checkmark}                               & {  \checkmark}                             & {  \checkmark}                                & {   \texttimes}                             \\
		\textbf{MCD}                       & {  \checkmark}                                  & {  \checkmark}                             & {  \checkmark}                               & {  \checkmark}                               & {  \checkmark}                             & {  \checkmark}                                & {  \checkmark} \\ \hline
	\end{tabular}
	\vspace{-5pt}
\end{table}

%\makecell{Training \\ Samples}

%What's more, the abnormal degree $h_{t,i}$ of MCD can not only be used as a defense method. The server of the FL system can use the abnormal degree $h_{t,i}$ of all clients in the CMD system to prioritize the clients, and then each round of client selection can be selected from clients with high priority. Select the local model that is most beneficial to the global model to maximize the training quality of the main task of the FL system. A more secure FL aggregation algorithm can be constructed based on this.

%\begin{figure}[h]
%	\centering
%	\includegraphics[width=\linewidth]{h}
%	\caption{abnormal degree $h_{t,i}$ in MCD under two attack scenarios}
%\end{figure}

\section{Conclusions and Future Work}

In this paper, we have proposed MCD, a defense countermeasure designed against stealthier data poisoning attacks in FL systems, which reviews local models accross multiple feature dimensions. To push the limit of MCD against GAN-based stealthier attacks, we have proposed VagueGAN, a GAN model specifically designed for data poisoning attacks against FL systems. Unlike traditional GANs, our VagueGAN has been verified to generate seemingly legitimate vague data with appropriate amounts of poisonous noise, and thus help strengthen MCD to work against stealthier GAN-based attacks as well as other mainstream ones. Our MCD has been validated to outperform existing defense methods against a range of data poisoning attacks, particularly stealthier GAN-based ones. 

%In this paper, we have proposed VagueGAN, a GAN model specifically designed for data poisoning attacks against FL systems. Unlike traditional GANs, our VagueGAN has been verified to generate seemingly legitimate vague data with appropriate amounts of poisonous noise. The ways to achieve a balanced trade-off between attack effectiveness and stealthiness have been studied. We have also experimentally demonstrated the distinctiveness of the poisoned data generated by VagueGAN compared to noisy data. Furthermore, we have suggested and verified MCD as a viable countermeasure to VagueGAN if the same GAN model is consistently applied by the attacker. 

In the future, two potential directions are worth pursuing. The first one involves further exploration of more threatening attack methods. In particular, the consistency of GAN outputs has been found to be effective in identifying GAN-poisoned data or models. Hence, a future study can develop more advanced generative models or adaptive attack strategies without such a drawback for improved attack stealthiness.  The second one focuses on exploring effective and universally applicable defense mechanisms. Although our MCD has been validated to be useful against GAN-based attacks, the model consistency it relies on may not hold if the attacks are implemented in varying settings, such as those involving adaptive attack strategies. Hence, a future study can produce better adaptive AI-driven techniques for enhanced defense capabilities. 

%In the future, we plan to extend our work in two potential directions. The first direction focuses on addressing the limitations of GAN-based attack methods. In particular, the consistency of GAN outputs has been found to be effective in identifying GAN-poisoned data or models. Hence, a future study can develop more advanced generative models without such a drawback for improved attack stealthiness. The second direction involves further exploration into effective and universally applicable defense mechanisms. Although our MCD has been validated to be useful against GAN-based attacks, the model consistency it relies on may not hold if the attacks are implemented in varying settings. Hence, a future study can produce better adaptive AI-driven techniques for enhanced defense capabilities.  

%\section*{Acknowledgment}

\bibliographystyle{IEEEtran}
\bibliography{paper}

\nocite{hitaj2017deep}
\nocite{chen2017targeted}
\nocite{shi2018spectrum}
\nocite{huang2021data}
\nocite{zhang2020online}
\nocite{schwarzschild2021just}
\nocite{tolpegin2020data}
\nocite{lyu2020threats}
\nocite{gosselin2022privacy}
\nocite{mothukuri2021survey}
\nocite{rodriguez2023survey}
\nocite{zhang2020poisongan}
\nocite{zhang2019poisoning}
\nocite{cao2019understanding}
\nocite{upreti2022defending}
\nocite{jere2020taxonomy}
\nocite{shen2016auror}
\nocite{li2021lomar}
\nocite{mu2024feddmc}
\nocite{goodfellow2020generative}
\nocite{shejwalkar2021manipulating}
\nocite{wang2020model}
\nocite{zhao2022detecting}
\nocite{li2019abnormal}
\nocite{mirza2014conditional}
\nocite{marfoq2021federated}
\nocite{xiao2017fashion}
\nocite{lecun1998mnist}
\nocite{krizhevsky2009learning}
\nocite{awan2021contra}
\nocite{zhang2022fldetector}
\nocite{cao2021provably}
\nocite{10054157}
\nocite{ovi2023confident}
\nocite{sikandar2023detailed}
\nocite{xia2023poisoning}
\nocite{lai2023two}
\nocite{gupta2023novel}
\nocite{rodriguez2023survey}
\nocite{yang2023clean}
\nocite{shen2023privacy}
\nocite{chen2016infogan}

\end{document}